\documentclass[twocolappendix]{emulateapj}
\usepackage{psfig,enumerate}
\usepackage{xcolor}
\usepackage{listings}
\usepackage{color}
\usepackage{atbegshi}
\usepackage{url}
\usepackage[caption=false]{subfig}

\begin{document}

\title{The stellar velocity distribution function in the Milky Way galaxy}

\shorttitle{Velocity DF in the solar vicinity}
\shortauthors{Anguiano et al.}

\author{
Borja Anguiano\altaffilmark{1}, Steven R. Majewski\altaffilmark{1}, Christian R. Hayes\altaffilmark{1}, Carlos Allende Prieto\altaffilmark{2,3}, Xinlun Cheng\altaffilmark{1}, Christian Moni Bidin\altaffilmark{4}, Rachael L. Beaton\altaffilmark{5,6,7}, Timothy C. Beers\altaffilmark{8}, and Dante Minniti\altaffilmark{9}}
\altaffiltext{1}{Department of Astronomy, University of Virginia, Charlottesville, VA, 22904, USA}
\altaffiltext{2}{Instituto de Astrof\'isica de Canarias (IAC), E-38205 La Laguna, Tenerife, Spain}
\altaffiltext{3}{Universidad de La Laguna, Dpto. Astrof\'isica, E-38206 La Laguna, Tenerife, Spain}
\altaffiltext{4}{Instituto de Astronom\'ia, Universidad Cat´olica del Norte, Av. Angamos 0610, Antofagasta, Chile}
\altaffiltext{5}{Department of Astrophysical Sciences, Princeton University, 4 Ivy Lane, Princeton, NJ 08544, USA}
\altaffiltext{6}{The Observatories of the Carnegie Institution for Science, 813 Santa Barbara St., Pasadena, CA 91101, USA}
\altaffiltext{7}{Hubble Fellow, Carnegie-Princeton Fellow}
\altaffiltext{8}{Department of Physics and JINA Center for the Evolution of the Elements, University of Notre Dame, Notre Dame, IN 46556, USA}
\altaffiltext{9}{Depto. de Cs. F\'isicas, Facultad de Ciencias Exactas, Universidad Andr\'es Bello, Av. Fern\'andez Concha 700, Las Condes, Santiago, Chile.}

\begin{abstract}

The stellar velocity distribution function (DF) in the solar vicinity
is re-examined using data from the SDSS APOGEE survey's DR16 and {\it Gaia} DR2. By exploiting APOGEE's ability to chemically discriminate with great reliability the thin disk, thick disk and (accreted) halo populations, we can, for the first time, derive the three-dimensional velocity DFs for these chemically-separated populations.

We employ this smaller, but more data-rich APOGEE+{\it Gaia} sample to build a \emph{data-driven model} of the local stellar population velocity DFs, and use these as basis vectors for assessing the relative density proportions of these populations over 5 $<$ $R$ $<$ 12 kpc, and $-1.5$ $<$ $z$ $<$ 2.5 kpc range as derived from the larger, more complete (i.e., all-sky, magnitude-limited) {\it Gaia} database. We find
that 81.9 $\pm$ 3.1$\%$ of the objects in the selected {\it Gaia} data-set are thin-disk stars, 16.6 $\pm$ 3.2$\%$ are thick-disk stars, and 1.5 $\pm$ 0.1$\%$ belong to the Milky Way stellar halo. We also find 
the local thick-to-thin-disk density normalization to be $\rho_{T}(R_{\odot})$/$\rho_{t}(R_{\odot})$ = 2.1 $\pm$ 0.2$\%$, a result consistent with, but determined in a completely different way than, typical starcount/density analyses.  Using the same methodology, the local halo-to-disk density normalization is found to be $\rho_{H}(R_{\odot})$/($\rho_{T}(R_{\odot})$ + $\rho_{t}(R_{\odot})$) = 1.2 $\pm$ 0.6$\%$, a value that may be inflated due to chemical overlap of halo and metal-weak thick disk stars. 

\end{abstract}



\section{Introduction}

Understanding galaxy formation is a major goal of current astrophysical research. Detailed studies of the stellar components of the Milky Way galaxy play a fundamental role in building this understanding 
because they provide direct and robust tests of and constraints on theories of galaxy formation. The latter are presently formulated within the framework of hierarchical formation 
in a $\Lambda$CDM Universe (e.g., \citealt{1991ApJ...379...52W,1998MNRAS.295..319M}), and the past few decades have seen a myriad of results affirming the role that mergers have had in the evolution of the Milky Way, no more vividly than the countless extant substructures that have been discovered and mapped throughout the Galactic stellar halo, whose origin appears to be dominated by the debris of these mergers.

Meanwhile, explorations of the evolution of the Galactic disk have a long and rich history that reaches back to the kinematical studies of
\cite{1915MNRAS..76...37E}, \cite{1915MNRAS..76...70J}, \cite{1920AJ.....33..113C}, \cite{1922ApJ....55..302K}, \cite{1926PhDT.........1O}, and \cite{1927ApJ....65..108S}, who explored the local stellar velocity distribution to gain information about large-scale Galactic structure.
The field significantly advanced 
when \cite{1950ApJ...112..554R} and \cite{1950AZh....27...41P} discovered a strong correlation between spectral type and UV-excess (i.e., metallicity) and the kinematical properties of stars near the Sun. 
One of the explanations of this phenomenon, which led to the seminal paper by \cite{1962ApJ...136..748E}, is that the oldest and least chemically enriched stars we observe today were born kinematically hot, and these turbulent birth velocities declined steadily as the gas in the Galaxy dissipated energy and settled into a dynamically cold disk
--- i.e., older populations were dynamically hotter \emph{ab initio} \citep[e.g.,][]{2004ApJ...612..894B,2013A&A...558A...9M}. 
However, it was later suggested that within some fraction of the thin disk, stars are born with random velocities that are small at birth but that increase with time (e.g., \citealt{1951ApJ...114..385S,1987ASIC..207..375F,2000MNRAS.318..658B,2004A&A...418..989N,2009MNRAS.397.1286A})
 
The velocity distribution function (DF) of stars in the Galaxy
can be a valuable aid in uncovering 
the relationships between kinematics, metallicity, and age for disk and halo stars, and lends insights into
the dynamical history of stellar populations
\citep{1999A&A...341...86B,2004MNRAS.350..627D,2005A&A...430..165F,2010ApJ...717..617B}.
In the Galactic stellar halo, where the timescales for dynamical mixing are long, the history of minor mergers imprints long-lived, dynamically cold substructures that are quite legible fossils from which Galactic archaeology can readily synthesize a sequence of events (e.g., \citealt{1994Natur.370..194I,2001ApJ...548L.165O,2003ApJ...599.1082M,2006ApJ...642L.137B,2006ApJ...643L..17G}).

However, a multitude of processes can affect the velocity DF in the Galactic disk, including the potential mixture of the aforementioned {\it ab initio} ``born hot'' and``born cold'' components, and 
with the latter affected by numerous \emph{disk heating} mechanisms.
The sum of these effects that increase velocity dispersions in disk stars complicate the simplest model of a spiral galaxy, where stars in the disk librate about circular orbits in the Galactic plane. From state-of-the-art magnetohydrodynamical 
cosmological simulations, \cite{2016MNRAS.459..199G} found that in the secular evolution of the disk, bar instabilities are a dominant heating mechanism. The authors also reported that heating by spiral arms \citep[e.g.,][]{2004MNRAS.350..627D,2006MNRAS.368..623M}, radial migration \citep[e.g.,][]{2002MNRAS.336..785S}, and adiabatic heating from mid-plane density growth \citep[e.g.,][]{1992MNRAS.257..620J} 
are all subdominant to bar heating. Another fundamental astrophysical process that scatters stars onto more eccentric orbits or onto orbits that are inclined to the disk's equatorial plane is the accretion of lower mass systems (see, e.g., \citealt{2008ApJ...688..254K,2014MNRAS.443.2452M}, and references therein). 
This is particularly relevant given that recent studies suggest that the Milky Way likely underwent such a merger $\sim$ 8 Gyr ago (e.g., \citealt{2010A&A...511L..10N,
2014ApJ...796...38N,
2017ApJ...842...49L,
2018MNRAS.478..611B, 2018MNRAS.tmp.1537K}). Moreover, \cite{2018MNRAS.481..286L}, using N-body simulations of the interaction of the Milky Way with a Sagittarius-like dSph galaxy, showed that the patterns in the large-scale velocity field reported from observations, like vertical waves (e.g., \citealt{2012ApJ...750L..41W}), can be described by tightly wound spirals and vertical corrugations excited by Sagittarius (Sgr) impacts. 

Studies of the density distribution of stars above and below the Galactic plane have shown that the Galactic disk is in fact composed of two distinct components \citep{1982PASJ...34..365Y,1983MNRAS.202.1025G}: a thin disk with a vertical scaleheight of 300 $\pm$ 50 pc, and a thick component with a scaleheight of 900 $\pm$ 180 pc at the distance of the Sun from the Galactic Center (see \citealt{2016ARA&A..54..529B} for a review of the spread in these quantities found by various authors using different methods). On the other hand,  
detailed stellar-abundance patterns for disk stars show a clear bimodal distribution in the 
[$\alpha$/Fe] versus [Fe/H] plane. This bimodality is associated with --- and firm evidence for --- a distinctly separated thin and thick disk
\citep{1993A&A...275..101E,2003A&A...410..527B,2011MNRAS.414.2893F,2013A&A...554A..44A}. 
The high--[$\alpha$/Fe] sequence, corresponding 
to the thick disk, exists over a large radial and vertical range of the Galactic disk (e.g., \citealt{2014ApJ...796...38N,2015ApJ...808..132H,2018MNRAS.476.5216D}). 

The existence of two chemically-distinguished disk components points to
a different chemical evolution and, hence distinct disk-formation mechanism and epoch (e.g.; \citealt{1997ApJ...477..765C,2015MNRAS.453.1855M,2019MNRAS.tmp..141C}) for the two disk components. Kinematically, the velocity dispersion for the chemically selected thick disk component is, on average, larger than the dispersion reported for the low--[$\alpha$/Fe] thin disk \citep{2008A&A...480...91S,2011ApJ...738..187L,2018MNRAS.474..854A}. 
However, there is substantial overlap in the thin and thick disk velocity DFs, which makes kinematics a less-reliable diagnostic for discriminating populations. From the standpoint of isolating and studying the chemodynamical properties of these populations free of cross-contamination, this is a bit unfortunate, because, despite many large, dedicated spectroscopic surveys of Galactic stars from which detailed chemical-abundance patterns can be measured for now millions of stars (see below), 
there are now several orders of magnitude more stars with {\it kinematical} data available, thanks to ESA's {\it Gaia} mission.

\emph{Gaia} is an all-sky astrometric satellite from which we can now obtain accurate sky position, parallax, and proper motion, along with the estimated uncertainties and correlations of these properties, for $\sim$ 1.3 billion sources. Moreover, the existence of a \emph{Gaia} sub-sample containing line-of-sight velocities provides unprecedented accuracy for individual space velocities for more than 7 million stellar objects \citep{2018arXiv180409365G}. Unfortunately, there are no $\alpha$-element abundance measurements for the vast majority of this \emph{Gaia} sub-sample. This prevents an unbiased and comprehensive study of the velocity DF for the thin disk, thick disk, and halo treated separately.

Fortunately, the astronomical community has invested 
great effort into building and developing massive multi-object spectroscopic surveys, where the measurement of abundances beyond the simple overall metallicity is possible. Projects like SEGUE \citep{2009AJ....137.4377Y}, RAVE \citep{2008AJ....136..421Z}, LAMOST \citep{2012RAA....12..723Z}, Gaia-ESO \citep{2012Msngr.147...25G}, GALAH \citep{2015MNRAS.449.2604D}, and APOGEE \citep{2017AJ....154...94M} are transforming our understanding of the Milky Way galaxy through their generation of vast chemical abundance databases on Milky Way stars.
The future is even more promising, 
with even larger spectroscopic surveys planned, such as
SDSS V \citep{2017arXiv171103234K}, WEAVE \citep{2012SPIE.8446E..0PD}, MOONS \citep{2016ASPC..507..109C}, DESI \citep{2016arXiv161100036D}, 4MOST \citep{2018IAUS..334..225F}, and MSE \citep{2016SPIE.9908E..1PZ}, all aiming to provide individual abundances for millions more 
Galactic stars in both hemispheres, together with line-of-sight velocities, $v_{\rm los}$, with a precision of a few hundred m s$^{-1}$. 
While the sum total of observed stars in these spectroscopic surveys will span only $\sim$1\% of even the present \emph{Gaia} sub-sample containing line-of-sight velocities, the chemical data in the former surveys can be used to begin characterizing and interpreting the vastly larger number of sources having kinematical data in the 
$\emph{Gaia}$ database.

One of the main goals of the present study is to perform, for the first time, a detailed and unbiased study of the Galactic velocity DFs --- derived from $\emph{Gaia}$ data --- for the individual, chemically separated stellar populations, and to explore how these distributions change for different Galactocentric radii and
distances from the Galactic mid-plane.
For this study we use the individual stellar abundances from the APOGEE survey, specifically [Mg/Fe] and [Fe/H], to relatively cleanly discriminate the thin and thick disks,
associated with the low and high $\alpha$-sequences, respectively, in the [$\alpha$/Fe]-[Fe/H] plane, as well as the halo stars, which predominantly inhabit other regions of the same plane.  Using the \emph{kinematical} properties of these \emph{chemically} defined sub-samples, we build a data-driven kinematical model, which we then apply to the full \emph{Gaia} database 
to ascertain the contribution of the different Galactic structural components to the velocity-space DF as a function of Galactic cylindrical coordinates, $R$ and $z$.  We also create two-dimensional maps in the $R$-$z$ plane, where we explore
the behavior of the thick-to-thin-disk density normalization and the halo-to-disk density normalization. 

This paper is organized as follows. In Section~\ref{APO_Gaia} we describe the APOGEE and \emph{Gaia} data-sets we employ in the analysis. Section~\ref{DF_disk} examines the velocity DF of the Galactic disk and halo, and describes the building of the data-driven model. We discuss the thick-disk normalization and the halo-to-disk density in Section~\ref{thick_norma}, and the most relevant results of the present study are summarized and discussed in Section~\ref{Conclusion}.

\section{The APOGEE-2 DR16 and \emph{Gaia} DR2 Data-sets}
\label{APO_Gaia}

Our study makes use of the data products from Data Release 16 (DR16) of the Apache Point Observatory Galactic Evolution Experiment (APOGEE, \citealt{2017AJ....154...94M}).  Part of both Sloan Digital Sky Survey III (SDSS-III, \citealt{2011AJ....142...72E}) and SDSS-IV \citep{2017AJ....154...28B} via APOGEE and APOGEE-2, respectively, the combined APOGEE enterprise has been in operation for nearly a decade, and, through the 
installation of spectrographs on both the Sloan \citep{2006AJ....131.2332G} and du Pont 2.5-m telescopes
\citep{1973ApOpt..12.1430B,2019arXiv190200928W},
has procured high-resolution, $H$-band spectra for more than a half million stars across both hemispheres.  The survey provides $v_{\rm los}$, 
stellar atmospheric parameters, and individual abundances for on the order of
fifteen chemical species \citep{2015AJ....150..148H,2015AJ....149..181Z,2016AJ....151..144G}. 
A description of the latest APOGEE data products from DR14 and DR16 can be found in \cite{2018AJ....156..125H} and J\"onsson et al. (in preparation), respectively.

In this study we also use 
data from the second data release (DR2) of the \emph{Gaia} mission 
\citep{2018arXiv180409365G}. This catalog provides full \emph{6-dimensional} space coordinates for 7,224,631 stars: positions ($\alpha$, $\delta$), parallaxes ($\varpi$), proper motions ($\mu^{*}_{\alpha}$, $\mu_{\delta}$), and radial line-of-sight velocities ($v_{\rm los}$) for stars as faint as $G$ = 13 \citep{2018arXiv180409369C}. The stars are distributed across
the full celestial sphere.  \emph{Gaia}  DR2 
contains $v_{\rm los}$ for stars with effective temperatures in the range $\sim$ 3550 - 6900 K. The median uncertainty for bright sources ($G < 14$) is 0.03 mas for the parallax and 0.07 mas yr$^{-1}$ for the proper motions \citep{2018arXiv180409366L,2018arXiv180409375A}. The precision of the \emph{Gaia} DR2 $v_{\rm los}$ at the bright end is on the order of 0.2 to 0.3 km s$^{-1}$, while at the faint end it is on the order of 1.4 km s$^{-1}$ for $T_{\rm eff}$ = 5000 K stars and $\sim$3.7 km s$^{-1}$ for $T_{\rm eff}$ = 6500 K. For further details about the \emph{Gaia} DR2 sub-sample containing $v_{\rm los}$ measurements, we refer the reader to \cite{2018arXiv180409371S} and \cite{2018arXiv180409372K}. We follow the recommendations from \cite{2019MNRAS.tmp..280B} for studies in Galactic dynamics using \emph{Gaia} to remove stars where the color photometry is suspect, as well as stars where the $v_{\rm los}$ measurement is based on fewer than four transits of the instrument.



Individual space velocities in a Cartesian Galactic system were
obtained by following the equations in
\citet{1987AJ.....93..864J}. That is, from \emph{Gaia} DR2
line-of-sight velocities, proper motions, and parallaxes, we derive
the space velocity components ($U$, $V$, $W$). In the case of the
APOGEE targets, we use the APOGEE-2 DR16 catalog $v_{\rm los}$
\citep{2019arXiv191202905A}, where the internal precision is better
than 0.1 km s$^{-1}$ \citep{2015AJ....150..173N}. When the relative
uncertainty in parallaxes become larger, the inverse of a measured
parallax is a biased estimate of the distance to a star
\citep{1927ApJ....65..108S,2018arXiv180409376L}. For this reason, we
select stars with positive parallaxes and a relative parallax
uncertainty smaller than 20$\%$ ($\sigma_{\varpi}/\varpi$ $\leq$ 0.2).
We also set the \emph{Gaia} flag astrometric-excess-noise = 0 to drop
stars with poor astrometric fits.  In addition, the flag rv-nb-transits > 4 is set to secure enough \emph{Gaia} transits that robust $v_{\rm los}$ measurements are in-hand.
Following \cite{2019MNRAS.tmp..280B}, the selected \emph{Gaia} stars for this study must have reported $G_{\rm BP}$ and $G_{\rm RP}$ magnitudes. That leaves 4,774,723 \emph{Gaia} targets for this exercise. The remaining
targets have high-precision \emph{Gaia} parallaxes (the vast majority of the stars have $\sigma_{\varpi}/\varpi$ $\leq$ 0.05), so that 
their distances can be determined by simple parallax inversion without significant biasing the error derivation (e.g., \citealt{2018AJ....156...58B} and references therein). We adopt a right-handed Galactic system, where $U$ is pointing towards the Galactic center, $V$ in the direction of rotation, and $W$ towards the North Galactic Pole (NGP). For the peculiar motion of the Sun, we adopt the values: $U_{\odot}$ = 11.1 km s$^{-1}$, $V_{\odot}$ = 12.2 km s$^{-1}$, and $W_{\odot}$ = 7.2 km s$^{-1}$
\citep{2010MNRAS.403.1829S}. We also transform the velocities from Cartesian to a cylindrical Galactic system: $\upsilon_{R}$, $\upsilon_{\phi}$, $\upsilon_{z}$. 
The Sun is assumed to be located at $X = 8.34$ $\pm$ 0.16 kpc and the circular rotation speed at the location of the Sun to be 240 $\pm$ 8 km s$^{-1}$ \citep{2014ApJ...783..130R}. We define $R = (X^{2} + Y^{2})^{1/2}$, as the distance from the Galactic center (GC), projected onto the Galactic plane. 

In the end, the typical uncertainty in the velocities used here is $\Delta\upsilon \sim$ 1.5 km s$^{-1}$ per dimension.
To remove outlier velocities that can yield unrealistic velocity dispersions we select stars where ($\upsilon_{R}^2 + (\upsilon_{\phi} - 240)^2 + \upsilon_{z}^2)^{1/2}$ $<$ 600 km s$^{-1}$, which removes a total of 503 stars. In the next section we explore the kinematical properties of the Milky Way using the data-set just described.

\section{The velocity distribution functions of the Galactic disk and halo populations}
\label{DF_disk}

Stars of different mass synthesize and, upon death, eject into the interstellar medium different chemical elements, and on different timescales. The overall metallicity measured in a star's atmosphere, for the most part, represents an integral over star formation and chemical enrichment prior to that star's birth, while the abundances of individual elements can be used to track the ratio of recent to past star-formation rates.

Thus, stellar chemical-abundance {\it patterns} offer a means to discriminate stellar populations that have experienced differing star-formation and chemical-enrichment histories. To procure unbiased velocity DFs for individual stellar populations in the nearby disk, we exploit the [Fe/H] and [Mg/Fe] abundances measured from APOGEE spectra, the combination of which has been show to give very good discrimination of the thin disk, thick disk, and halo (e.g., \citealt{2003A&A...410..527B,2018ApJ...852...49H,2019ApJ...874..102W}). We select from the APOGEE database those 
stars for which S/N $>$ 70 and no aspcapbad flag is set \citep{2015AJ....150..148H}. The APOGEE survey has a number of focused science programs that target specific objects
like the Sgr dSph galaxy \citep{2013ApJ...777L..13M,2017ApJ...845..162H}, 
the Large and Small Magellanic Clouds \citep{2019arXiv190103448N}, numerous star clusters \citep{2020MNRAS.492.1641M}, 
and many other stellar astrophysics programs \citep{2013AJ....146...81Z,2017AJ....154..198Z} 
that are not germane to this study of {\it normal field stars}.
To remove these specialized targets from our database, we identified fields associated with the special programs, and removed all targets within those fields.

\begin{figure}[ht]
\begin{center}
\includegraphics[width=1.06\hsize,angle=0]{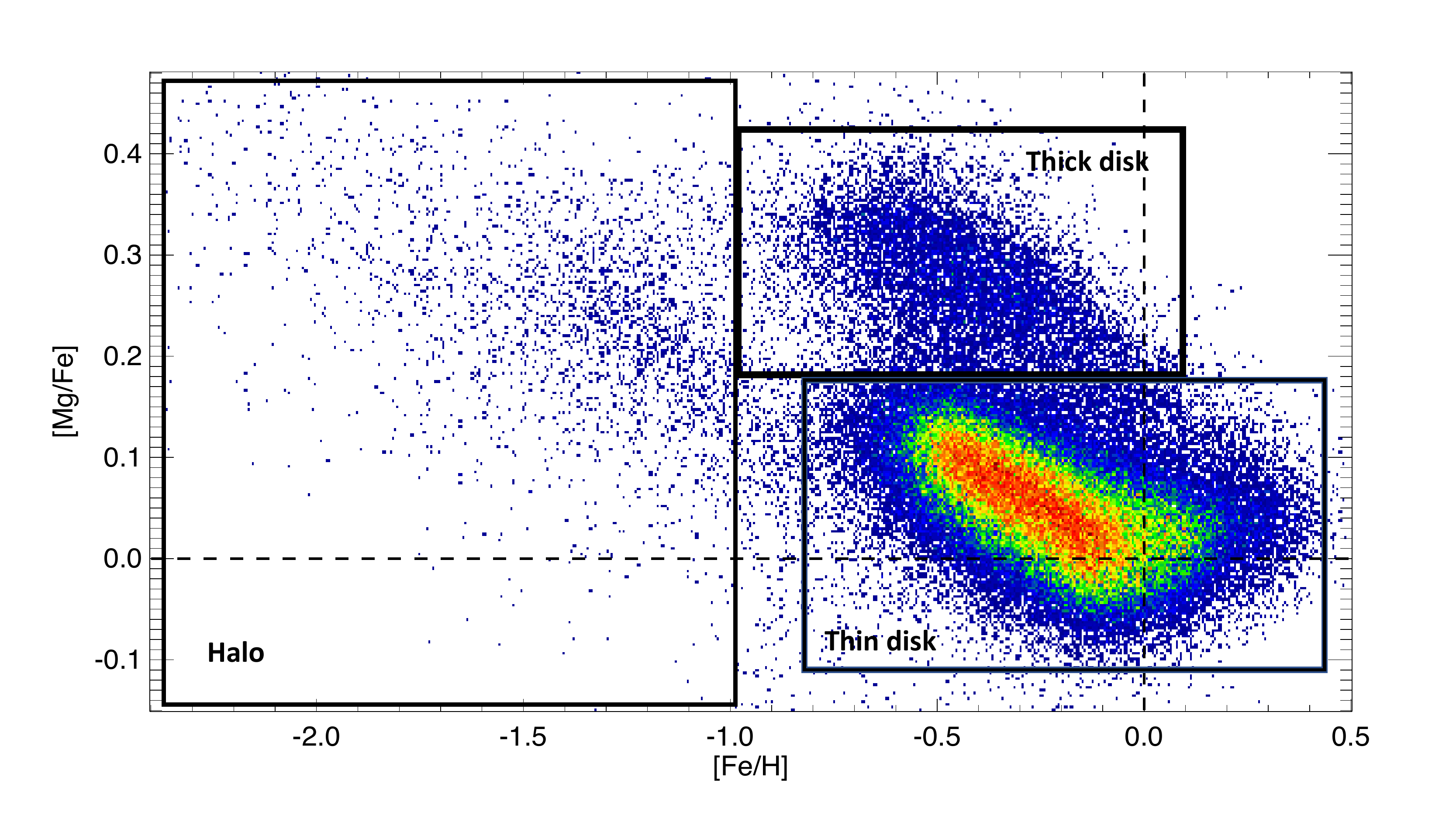}
\end{center}
\caption{The distribution of [Mg/Fe] with [Fe/H] abundances for our curated sample of stars from the APOGEE survey. We use these abundances to select stars with low-$\alpha$ abundances as thin disk, and those with high-$\alpha$ abundances as stars associated with the thick disk, as indicated by the selection boxes. For the halo we select all the stars with [Fe/H] $\leq$ $-1.0$. The color code indicates stellar density, where red shows the highest density of stars and  blue the lowest. The dashed lines show the solar values for reference.}
\label{abund_APOGEE}
\end{figure}

\begin{figure*}[ht]
\begin{center}
\includegraphics[width=1\hsize,angle=0]{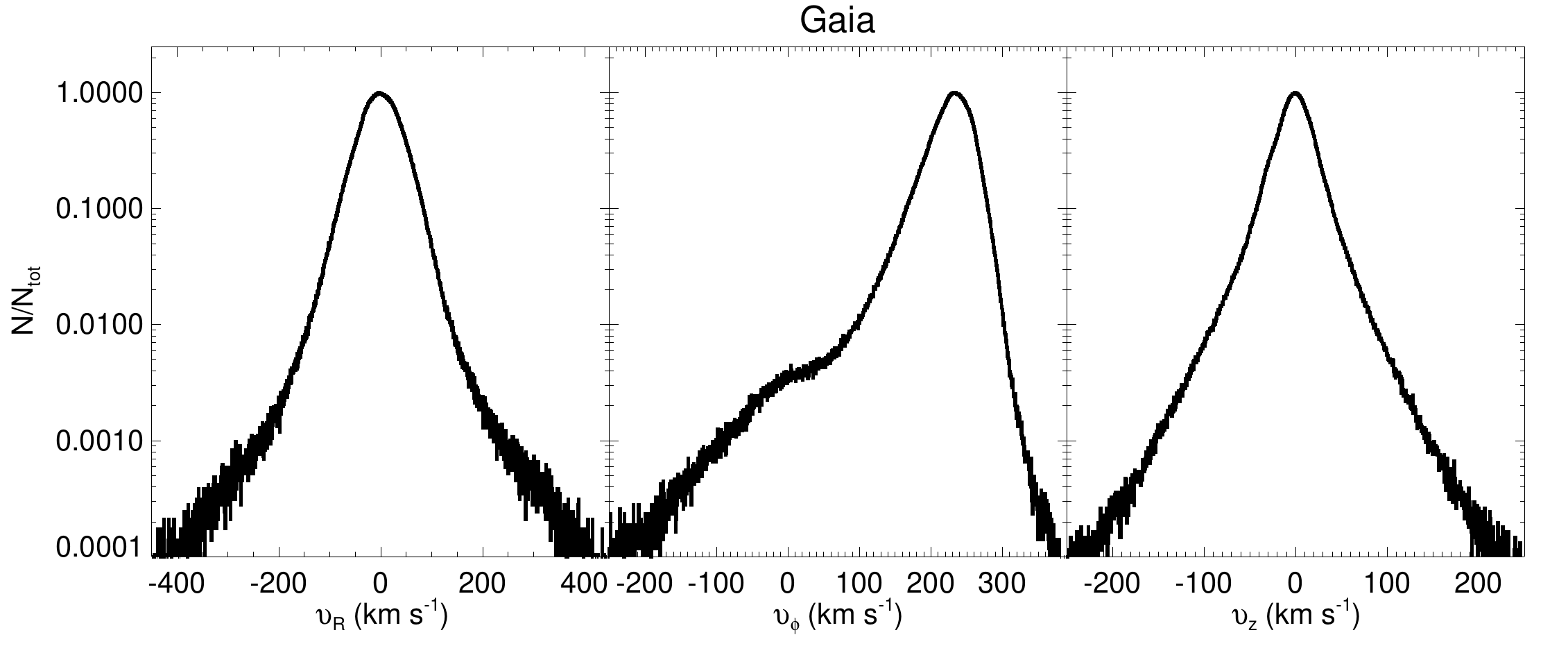}
\includegraphics[width=1\hsize,angle=0]{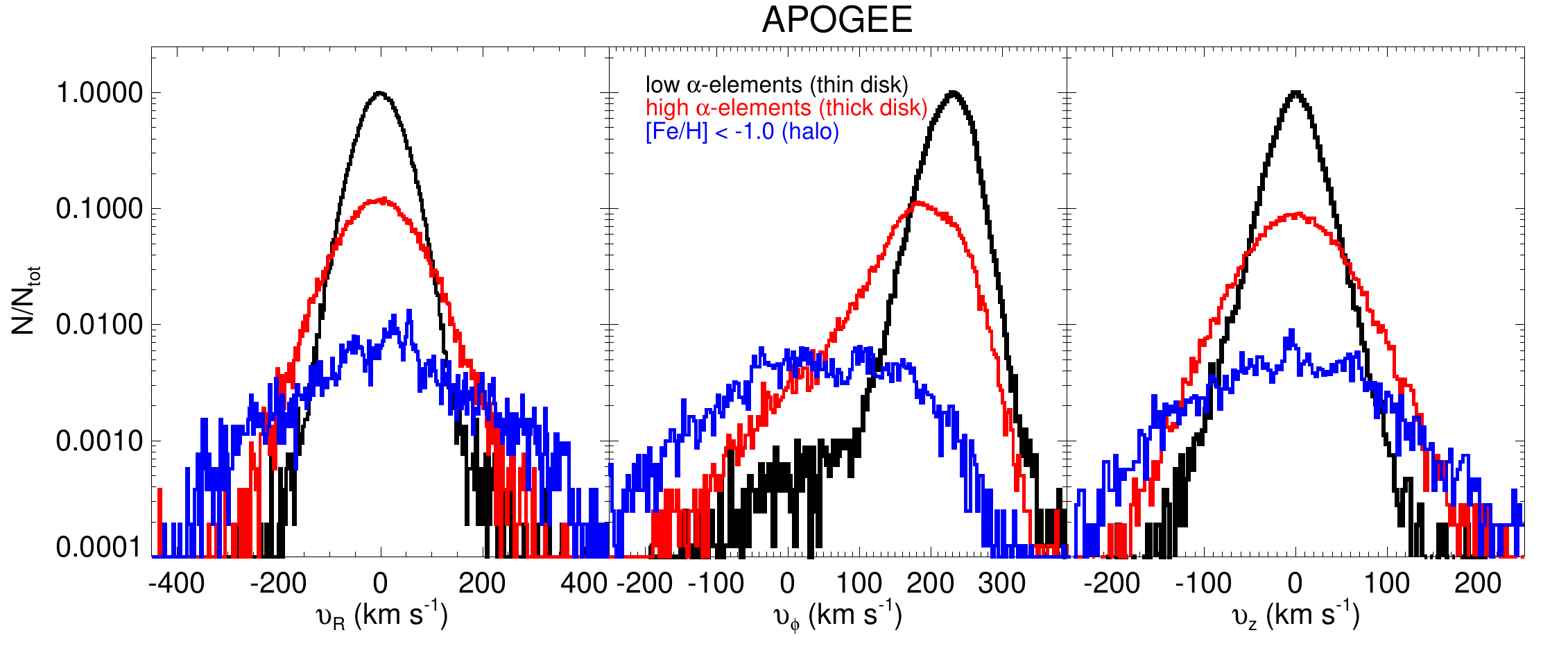}
\end{center}
\caption{The
relative distribution functions (normalized to a peak value of unity) for individual velocity components on a logarithmic scale using the \emph{Gaia} data-set (upper panels) and the APOGEE survey sample (lower panels). For the APOGEE sample we can assign stars to their respective stellar populations (indicated by different colors) using their chemistry, 
to reveal the kinematical properties of each population with little cross-contamination from the others.
}
\label{vphi_APOGEE}
\end{figure*}

We summarize our chemistry-based discrimination and selection of our three primary stellar populations  
in Figure~\ref{abund_APOGEE}. 
The higher [Fe/H] population with low-$\alpha$ abundances (--0.7 $<$ [Fe/H] $<$ +0.5, --0.1 $<$ [Mg/Fe] $<$ +0.17) is associated with the Galactic \emph{thin disk}, and our selection contains a total of 211,820 of these stars. The [Fe/H] $<$ 0 stars with high-$\alpha$ abundances ($+0.18$ $<$ [Mg/Fe] $<$ $+0.4$ within the metallicity range $-1.0$ $<$ [Fe/H] $<$ 0.0) we associate with the Galactic \emph{thick disk}, and the number of stars we have in that sample is 52,709. Finally, the \emph{Galactic halo} population is identified with stars having [Fe/H] $<$ --1.0 (e.g., \citealt{2018ApJ...852...49H}). 
The number of APOGEE stars in this population is 5,795. The magnitude-limited \emph{Gaia} data-set used in this study contains different stellar populations associated with different structures in the Galaxy. This is evident in Figure~\ref{vphi_APOGEE}, where we show the distribution of $\upsilon_{R}$, $\upsilon_{\phi}$, and $\upsilon_{z}$ for the population-integrated \emph{Gaia} sample and for the population-separated APOGEE-2 data.  
From a comparison of the upper and lower panels it becomes obvious that an assignment of population membership to a particular star
based exclusively on kinematical properties is not generally possible.
Moreover, it is not possible on the basis of kinematical data alone to determine with reliability even the relative contributions of the different populations to the net velocity DF on a statistical basis.
Figure~\ref{vphi_APOGEE} shows 
that the velocity DF of the different Galactic components 
clearly overlap, but also 
that individual abundances from high-resolution spectroscopy surveys are a useful tool for apportioning stars to their relative stellar populations
\citep{2011MNRAS.412.1203N,2014ApJ...796...38N}. 

%
While severe overlap is expected in the $\upsilon_{R}$ and $\upsilon_{z}$ dimensions, where all three stellar populations share a mean values around 0 km s$^{-1}$, we expect more separation of populations in $\upsilon_{\phi}$ due to variations in asymmetric drift. Nevertheless, stars with $\upsilon_{\phi}$ $>$ 70 km s$^{-1}$ dominate the distribution, and yet have contributions from all three populations, though, naturally, most strongly from the thin and thick disk. We also observe that the thin-disk population shows a very small number of slow-rotating stars; these could represent a small contaminating portion of stars from the
accreted halo, which can have metallicities as high as [Fe/H] $\sim$ $-0.5$ \citep{2018ApJ...852...49H}.

Interestingly, for the population with [Fe/H] $<$ $-1.0$, we observe an extended velocity DF with a peak around $\upsilon_{\phi}$ $\sim$ 0 km s$^{-1}$, together with a second peak around 120 km s$^{-1}$ (see the middle panel in Figure~\ref{vphi_APOGEE}). We discuss this metal-poor population in more detail in Section~\ref{halo_pop}. 




While there is no precedent to the accuracy and precision of the \emph{Gaia} astrometry, and for such an enormous number of objects, there is still no detailed chemistry available for the 
vast majority of the stars in the $Gaia$ data-set.
Thus, one cannot yet leverage the huge statistical power of \emph{Gaia} to 
explore the kinematical properties of individual stellar populations in an unbiased manner, nor use these kinematics to sort stars reliably into populations to help define other gross properties of the populations.  However, we will now show how, with a \emph{data-driven model} trained with the velocity DFs defined for the combined APOGEE+\emph{Gaia} dataset for relatively nearby stars, we can harness the power of the greater \emph{Gaia} database without chemical data to sort stars into populations based on their kinematics, and use these statistical memberships to ascertain other bulk properties of the populations --- e.g., to 
assess the relative densities of stars in these populations over a broad range of the Galactic locations.
%
We describe the steps toward these goals in the next sections.  

\begin{figure*}[ht]
\begin{center}
\includegraphics[width=1.\hsize,angle=0]{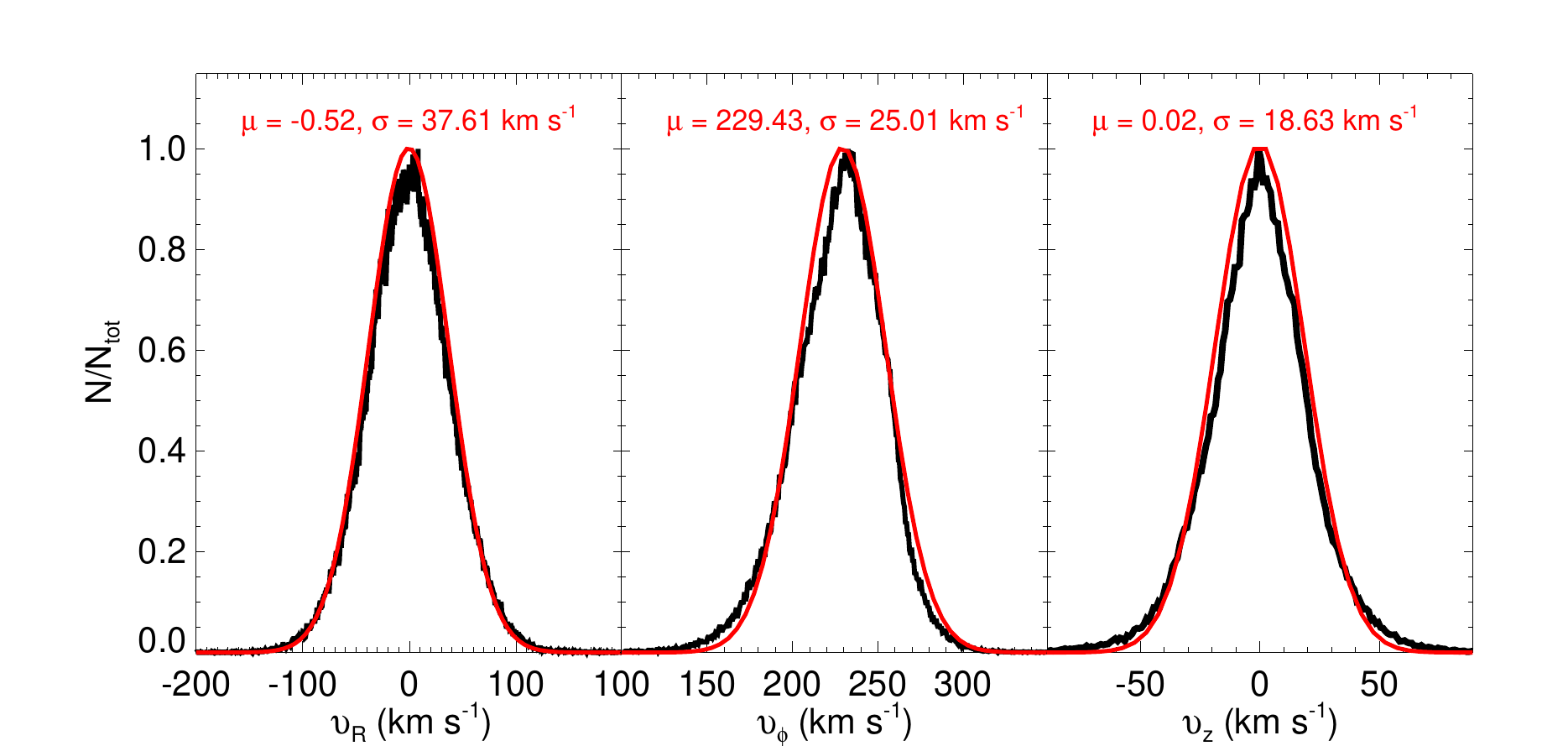}
\end{center}
\caption{Velocity distribution function for the chemically selected APOGEE thin disk (black distribution) for the $\upsilon_{R}$, $\upsilon_{\phi}$ and $\upsilon_{z}$ velocity components, respectively. The red distributions in the three panels show the best-fit following a single Gaussian function. The mean and the standard deviation of the normal distributions are shown in each panel.}
\label{velo_thin_apogee}
\end{figure*}

\subsection{APOGEE Data-Driven Model}
\label{model}

To build a simple, kinematical data-driven model we do not assume a triaxial Gaussian distribution function \citep{1907NWGot...5..614S,1999A&A...341...86B}; instead, we use the velocity DF for the chemically selected thin-disk, thick-disk and halo Galactic components in the APOGEE data. The DF, $f(\vec{\upsilon}$), is defined such that $f(\vec{\upsilon}$) d$\vec{\upsilon}$ is the number of stars per unit volume with velocity in the range [$\vec{\upsilon}$,$\vec{\upsilon}$ + d$\vec{\upsilon}$]. We now explore the characteristics of the DF for each primary Galactic stellar population.


\begin{table*}
\caption{Summary of the kinematical properties for different populations subsamples 
}
\begin{center}
\begin{tabular}{ccccccc}
\hline
& $\overline{\upsilon_\phi}$ & $\sigma_{\rm R}$ & $\sigma_{\rm \phi}$  &  $\sigma_{\rm  z}$ & $\sigma_{\phi}$/$\sigma_{\rm R}$ & $\sigma_{z}$/$\sigma_{\rm R}$ \\
& (km s$^{-1}$) & (km s$^{-1}$) & (km s$^{-1}$) & (km s$^{-1}$) & & \\
 \hline
 \hline
Chemically selected thin disk & +229.43 $\pm$ 0.54 & 37.61 $\pm$ 0.07 & 25.01 $\pm$ 0.04 & 18.53 $\pm$ 0.03 & 0.66 $\pm$ 0.01 & 0.49 $\pm$ 0.01 \\
Chemically selected thick disk & +191.82 $\pm$ 0.24 & 64.68 $\pm$ 0.20 & 50.82 $\pm$ 0.15 & 43.60 $\pm$ 0.13 & 0.78 $\pm$ 0.01 & 0.67 $\pm$ 0.01 \\ 
Halo ([Fe/H] $<$ -1) & +35.53 $\pm$ 1.28 & 150.57 $\pm$ 1.58 & 115.67 $\pm$ 1.21 & 86.67 $\pm$ 0.91 & 0.77 $\pm$ 0.02 & 0.57 $\pm$ 0.01 \\
\hline
Disk-like rotation & +186.10 $\pm$ 1.66 & 59.23 $\pm$ 1.39 & 50.00 $\pm$ 1.17 & 47.55 $\pm$ 0.85 & 0.84 $\pm$ 0.03 & 0.80 $\pm$ 0.02 \\
Non-rotation & --2.35 $\pm$ 1.57 & 165.51 $\pm$ 1.94 & 95.12 $\pm$ 1.11 & 94.10 $\pm$ 1.16 & 0.57 $\pm$ 0.01 & 0.57 $\pm$ 0.01 \\

\end{tabular}
\end{center}
\label{Tab1}
\end{table*}

\subsubsection{Chemically Selected Thin Disk}
\label{thin}

Figure~\ref{velo_thin_apogee} shows the velocity DF for $\upsilon_{R}$, $\upsilon_{\phi}$, and $\upsilon_{z}$ of the low-$\alpha$ sequence population selected in the [Fe/H]-[Mg/Fe] plane (see Figure~\ref{abund_APOGEE}). We also show the best Gaussian fit for each of the three components of velocity (red lines in Figure~\ref{velo_thin_apogee}). 

By and large, these Gaussian fits are reasonable descriptors of the DFs.  Even for the $\upsilon_{\phi}$ component, where we expect a skew in the observed DF due to 
asymmetric-drift effects, 
a normal distribution nevertheless reproduces the distribution reasonably well. We find the largest discrepancies between the observed velocity DF and a Gaussian distribution in all three cases to be mainly 
at the very peaks and in the wings, and, for the latter, especially in the cases of $\upsilon_{\phi}$ and $\upsilon_{z}$. As a demonstration of the utility of Gaussians as descriptors of the DFs, we find that for the chemically selected thin disk the DF skewness($\upsilon_{R}$)\footnote{Because there is no standard naming convention for the variables skewness and kurtosis, and some of the adopted variable names in the literature are redundant with those we use for other quantities, we simply employ the variable names ``skewness'' and ``kurtosis'' here to avoid confusion.} = 0.04, and the kurtosis($\upsilon_{R}$) = 0.61. For the azimuthal velocity, we have skewness($\upsilon_{\phi}$) = --0.44, and kurtosis($\upsilon_{\phi}$) = 0.98. Furthermore, for the vertical component of the velocity, we find skewness($\upsilon_{z}$) = --0.06 and a kurtosis($\upsilon_{z}$) = 1.33. These skewness and kurtosis values lie within the 
allowable range for normal univariate  distributions; e.g., \cite{DM10} argue that values for asymmetry and kurtosis between --2 and +2 are indicative
of normal univariate distributions, while even by the more conservative limits of --1.5 and +1.5 advocated by \cite{2012MNRAS.425..969P}, the Figure~\ref{velo_thin_apogee} DFs are found to be well-described as Gaussians. 

For the low-$\alpha$ sequence population we find the following velocity dispersion values, ($\sigma_{R}$, $\sigma_{\phi}$, $\sigma_{z}$) = (36.81 $\pm$ 0.07, 24.35 $\pm$ 0.04, 18.03 $\pm$ 0.03) km s$^{-1}$.  
Based on these values, the shape of the velocity ellipsoid for the thin disk is found to be ($\sigma_{R}:\sigma_{\phi}:\sigma_{z}$) =  (1.00:0.66:0.49).  
We also find that the vertex deviation for this population to be $\alpha_{R\phi}$ = --4.01$^{\circ}$ $\pm$ 0.09$^{\circ}$, and the tilt of the velocity ellipsoid to be $\alpha_{Rz}$ = +1.41$^{\circ}$ $\pm$ 0.02$^{\circ}$; these quantities are defined so that $\alpha_{ij}$ corresponds to the angle between the $i$-axis and the major axis of the ellipse formed by projecting the three-dimensional velocity ellipsoid onto the $ij$-plane, where $i$ and $j$ are any of the stellar velocities. We can also calculate the orbital anisotropy parameter \cite{1980MNRAS.190..873B} in spherical polar coordinates; for the thin disk we have $\beta$ = +0.57 $\pm$ 0.01.  These findings are summarized in Table~\ref{Tab1} and Table~\ref{Tab2}.  
%

\begin{figure*}[ht]
\begin{center}
\includegraphics[width=1.\hsize,angle=0]{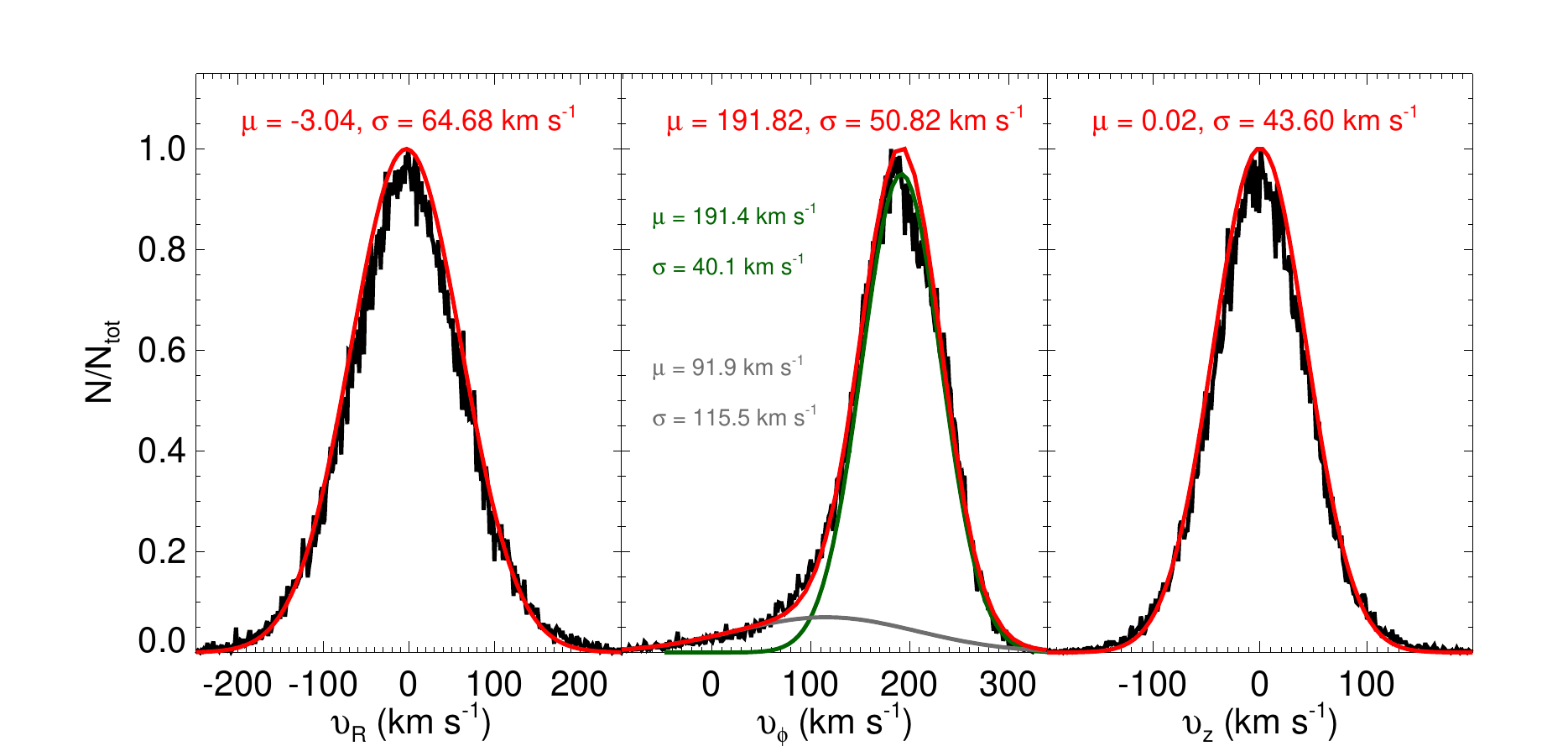}
\end{center}
\caption{Same as Figure \ref{velo_thin_apogee}, but for the thick disk.
However, for the $\upsilon_{\phi}$ component the red curve represents the sum of a two-component Gaussian decomposition of the distribution, needed to account for the asymmetric tail due to a broader spread in asymmetric drift (see text).  The components of this two-component fit are shown by the green and grey curves and listed values.
}
\label{velo_thick_apogee}
\end{figure*}

\subsubsection{Chemically Selected Thick Disk}
\label{thick}

In Figure~\ref{velo_thick_apogee}, the velocity DF for the three individual space velocities is shown for the high-$\alpha$ sequence population defined in Figure~\ref{abund_APOGEE}.  For this population we find a mean rotational velocity of $\overline{\upsilon_{\phi}}$ = 191.82 $\pm$ 0.24 km s$^{-1}$ and velocity dispersion of  ($\sigma_{R}$, $\sigma_{\phi}$, $\sigma_{z}$) = (62.44 $\pm$ 0.21, 44.95 $\pm$ 0.15, 41.45 $\pm$ 0.15) km s$^{-1}$.  As in the case for the thin disk, we find that normal distributions (red lines in Figure~\ref{velo_thick_apogee}) give a good match for the velocity components $\upsilon_{R}$ and $\upsilon_{z}$. For example, we calculate a skewness = 0.03, and kurtosis = 0.62 for the radial velocity component, while for the vertical velocity we have 0.02 and 0.86, respectively.

However, for the rotational velocity ($\upsilon_{\phi}$) of the thick-disk population the spread in asymmetric-drift is more prominent than for the thin disk,
so that the 
$\upsilon_{\phi}$ DF is 
skewed to low velocities. For this velocity component we have a skewness = $-1.38$ and a kurtosis = $-4.46$.  
Nevertheless, remarkably, it is possible to account for this skewness adequately
with only a single other
normal distribution (gray line) fit to 
the $\upsilon_{\phi}$ DF (see middle panel in Figure~\ref{velo_thick_apogee}). 
While the simplicity of a two-Gaussian fit is very convenient, it is not clear that it has a physical meaning as representing two distinct sub-populations, or that the mathematical contrivance just happens to work well as an explanation for what could be a more complex combination of sub-populations. If it really has a physical meaning as an apparent second sub-population in $\upsilon_{\phi}$, this less-dominant sub-population has Gaussian parameters $\mu$ $\sim$ 92 km s$^{-1}$ and $\sigma$ $\sim$ 115 km s$^{-1}$, and could represent the oldest population in the thick disk, which is, on average, lagging some 100 km s$^{-1}$ behind a more rapidly rotating, and significantly dynamically colder and presumably younger thick-disk population (green distribution in Figure~\ref{velo_thick_apogee}). 

We find differences in the shape of the velocity ellipsoid for the thick disk with respect to the thin disk (Table~\ref{Tab1}), where ($\sigma_{R}:\sigma_{\phi}:\sigma_{z}$) =  (1.00:0.78:0.67). For purposes of this calculation, we treat the total thick-disk sample as one population; this is certain to result in some increase in the measured dispersion in the $\upsilon_{\phi}$ direction.

Theoretical studies in the formation of the thick disk predict a wide range of values for $\sigma_{z}$/$\sigma_{R}$. For example, \cite{2010ApJ...718..314V} found a range from $\sim$ 0.4 to 0.9 for a model with formation of the thick disk through heating due to accretion events. Assuming that a merger led to the dynamical heating of a pre-existing precursor disk 
to the thick disk (e.g., \citealt{2018ApJ...863..113H,2019A&A...632A...4D,2019NatAs...3..932G} and references therein), our findings may be suggestive of an encounter with a satellite on a low/intermediate orbital inclination. 

Interestingly, we find the tilt of the velocity ellipsoid for the thick disk to be larger than that found for the thin disk, however, our tilt $\alpha_{Rz}$ = +5.16$^{\circ}$ $\pm$ 0.15$^{\circ}$, is lower than previous values reported for the thick disk. For example, \cite{2011ApJ...728....7C} found $\alpha_{Rz}$ = +8.6$^{\circ}$ $\pm$ 1.8$^{\circ}$, where the authors selected the thick disk population using stellar density laws and the vertical height with respect to the Galactic plane. With a sample of $\sim$1200 red giants, \citet{2012ApJ...747..101M} found an even larger value, $\alpha_{Rz}$ = +10.0$^{\circ}$ $\pm$ 0.5$^{\circ}$. The latter authors created the thick disk sample by selecting stars with $\vert$z$\vert$ $>$ 1.3 kpc, and they use the individual velocities of each star to remove outliers they associate with the halo. However, a direct comparison of our results 
with previous measurements from the literature is difficult, because the tilt angle varies with $z$ \citep{2012ApJ...747..101M,2020MNRAS.493.2952H}.

We also find the orbital anisotropy parameter for the Galactic thick disk, $\beta$ = +0.32 $\pm$ 0.01, to be lower than the value found for the thin disk, which suggests that the orbital anisotropy is mildly radial for this population. 

\begin{figure*}[ht]
\begin{center}
\includegraphics[width=1.\hsize,angle=0]{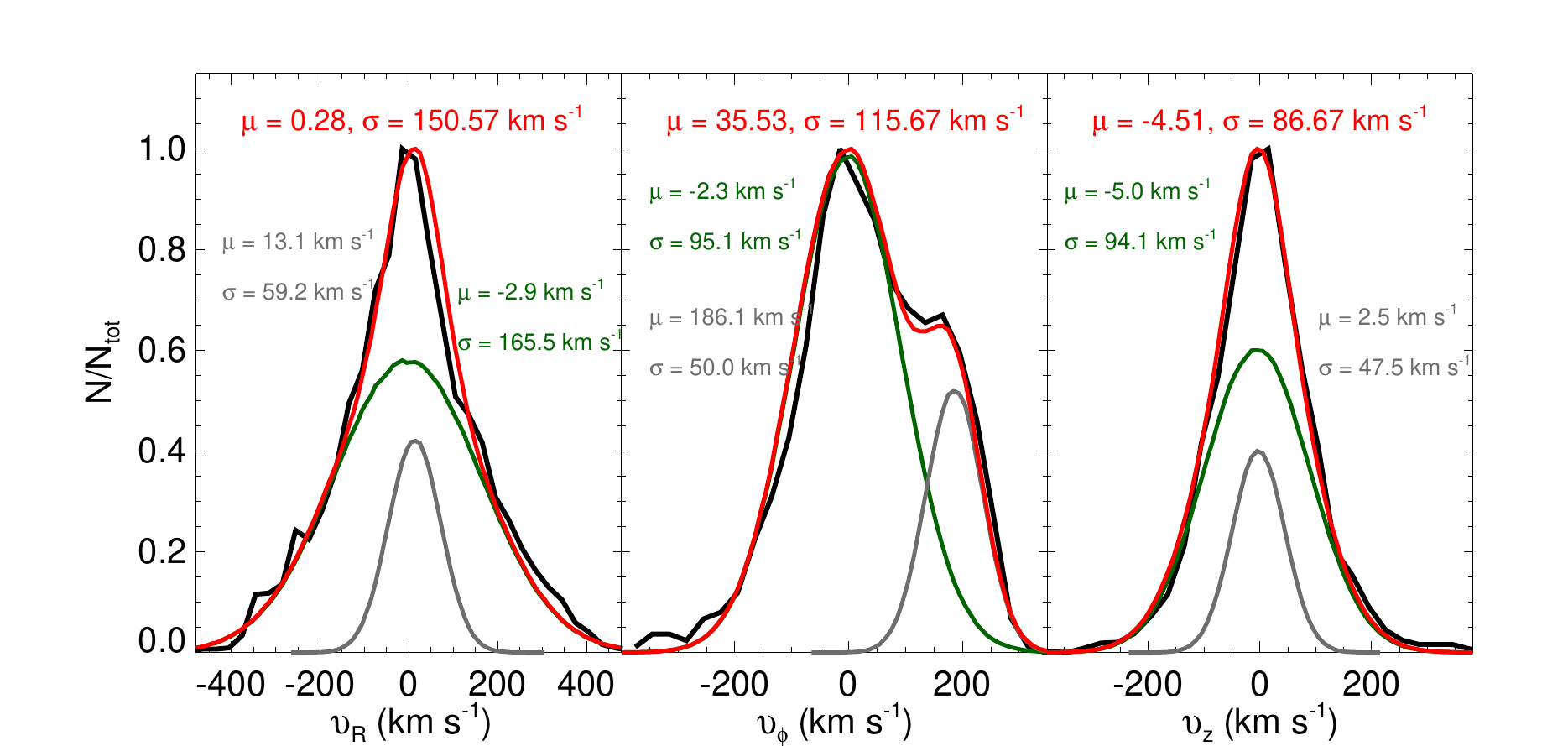}
\end{center}
\caption{Same as Figures \ref{velo_thin_apogee} and \ref{velo_thick_apogee}, but for the Milky Way halo, defined by stars with [Fe/H] $<$ $-1.0$ (black distribution).
In this case, each of the three velocity components are fitted by 
a mixture of two Gaussian components (see text for details). The mean and the standard deviation of the mixture distributions are shown in the top of each panel in red, and for each Gaussian component in grey and green.}
\label{velo_halo_apogee}
\end{figure*}

\begin{table*}
\caption{Tilt of the velocity ellipsoid, vertex deviation, and orbital anisotropy parameter for different populations}
\begin{center}
\begin{tabular}{cccc}
\hline
 & $\alpha_{\rm Rz}$ & $\alpha_{\rm R\phi}$ & $\beta$ \\
 & (degrees) & (degrees) & \\ 
 \hline
 \hline
Chemically selected thin disk & +1.41 $\pm$ 0.02 & --4.01 $\pm$ 0.09 & +0.57 $\pm$ 0.01 \\
Chemically selected thick disk & +5.16 $\pm$ 0.15  & --6.01 $\pm$ 0.47 & +0.32 $\pm$ 0.01 \\
Halo ([Fe/H] $<$ -1) & +11.43 $\pm$ 0.35  & --2.30 $\pm$ 0.15 & +0.20 $\pm$ 0.02 \\
\hline
Disk-like rotation & +0.86 $\pm$ 0.38 & +1.18 $\pm$ 0.59 & +0.56 $\pm$ 0.02 \\ 
Non-rotation & +0.15 $\pm$ 0.02 & +0.16 $\pm$ 0.02 & +0.42 $\pm$ 0.02 \\
\end{tabular}
\end{center}
\label{Tab2}
\end{table*}

\subsubsection{Halo}
\label{halo_pop}

The stellar density of halo stars, simply defined here by stars having [Fe/H] $<$ $-1.0$, is significantly lower than that for the disk-like populations in our volume of study, hence, the selection effects driven by the APOGEE survey design \citep{2017AJ....154..198Z} may have a larger impact on this population. The black lines in Figure~\ref{velo_halo_apogee} show the velocity DF for the halo population. For the distribution in $\upsilon_{\phi}$, we use a mixture of two Gaussian components. 
The metal-poor stars with disk-like kinematics (gray Gaussian in the middle panel of Figure~\ref{velo_halo_apogee}) may be associated with the thick disk, while the components with a non-rotation average motion may be associated with the inner halo (see \citealt{2000AJ....119.2843C}, \citealt{2010ApJ...712..692C}, \citealt{2018ApJ...852...49H}, \citealt{2018MNRAS.tmp.2814M} and references therein for more details). 
The stars in the entire halo sample ([Fe/H] $<$ $-1.0$)
are characterized by a radially elongated velocity ellipsoid, where we have ($\sigma_{R}$, $\sigma_{\phi}$, $\sigma_{z}$) = (150.57 $\pm$ 1.52, 115.67 $\pm$ 1.08, 86.67 $\pm$ 0.93) km s$^{-1}$. For the entire halo population defined in this study (red lines in Figure ~\ref{velo_halo_apogee}), we find a small mean prograde rotation of 35 km s$^{-1}$. These results are in good agreement with the halo properties reported in \cite{2000AJ....119.2843C} and \cite{2010ApJ...716....1B},
who also defined the halo by simple metallicity cuts.
Finally, for the entire [Fe/H] $<-1.0$ halo sample we find a large angle for the tilt of the velocity ellipsoid, 
$\alpha_{Rz}$ = $+11.43^{\circ}$ $\pm$ $0.35^{\circ}$, while $\beta$ is found to be nearly isotropic for these stars.
  

Meanwhile, the ``halo'' sub-population with disk-like kinematics in our sample shows a mean rotational velocity of $\overline{\upsilon_{\phi}}$ = 186.10 $\pm$ 1.36 km s$^{-1}$ (gray Gaussian in Figure~\ref{velo_halo_apogee}); this velocity is very similar to the mean rotational velocity we found for the Galactic thick disk (see the middle panel in Figure~\ref{velo_thick_apogee}). For the velocity dispersion of this sub-population, we find ($\sigma_{R}$, $\sigma_{\phi}$, $\sigma_{z}$) = (59.25 $\pm$ 1.12, 50.00 $\pm$ 0.91, 47.52 $\pm$ 0.94) km s$^{-1}$. These results are consistent with the values reported for the chemically selected thick disk (see Sect.~\ref{thick}), supporting previous results showing that the thick disk might exhibit an extended metal-poor tail --- more metal-poor than [Fe/H] = $-1.0$ and reaching values of [Fe/H] $\sim$ $-1.5$, or even lower (i.e.,  \citealt{1986ApJS...61..667N,1990AJ....100.1191M,2002AJ....124..931B,2010A&A...511L..10N,2014A&A...569A..13R,2018ApJ...852...49H,2018ApJ...852...50F,2019arXiv191206847A}).

For the other, ``more traditional'' halo component with a non-rotating average motion of  $\overline{\upsilon_{\phi}}$ = -2.31 $\pm$ 1.59 km s$^{-1}$ (green Gaussian in Figure~\ref{velo_halo_apogee}), we find that the velocity dispersion is ($\sigma_{R}$, $\sigma_{\phi}$, $\sigma_{z}$) = (165.52 $\pm$ 1.97, 95.10 $\pm$ 1.12, 94.14 $\pm$ 1.14) km s$^{-1}$. This halo population has been extensively discussed in the literature (e.g., \citealt{1978ApJ...225..357S,1985ApJ...291..260R,1998AJ....116..748G,2003A&A...406..131G,2019A&A...632A...4D}, and references therein).

These properties for the hotter halo component are consistent with current descriptions of the halo as an accreted population of the Milky Way.
For example, \cite{2010A&A...511L..10N} used $\sim$100 halo stars in the solar neighborhood to identify a 'low-$\alpha$' population, for which the kinematics 
suggest that it may have been accreted from dwarf galaxies \citep{2009ApJ...702.1058Z,2010MNRAS.404.1711P}, some specifically
originating from the $\omega$ Cen progenitor galaxy \citep{2003MNRAS.346L..11B,2005MNRAS.359...93M,2012ApJ...747L..37M}. Meanwhile, \cite{2018MNRAS.478..611B} studied the orbital anisotropy for the local stellar halo, and, by comparing the observational results with cosmological simulations of halo formation, concluded that the inner halo was deposited in a major accretion event by a satellite with $M_{\rm vir}$ $>$ 10$^{10}$ M$_{\odot}$, inconsistent with a continuous accretion of dwarf satellites.
\cite{2018MNRAS.478..611B} also highlighted the non-asymmetric structure of the remains of the merger in the velocity distribution.
%
Their findings 
echo discussions presented in  simulations including a similar accretion event by \cite{2005MNRAS.359...93M}, as well as the exploration of the \cite{2004AJ....128.1177V} data-set explored by \cite{2011MNRAS.412.1203N}, where the locally sampled velocity distribution of a high-energy accretion event appears to produce a mixture of kinematical populations.

In addition to the \cite{2018MNRAS.478..611B} study, \cite{2018Natur.563...85H} selected the retrograde halo population using APOGEE DR14 \citep{2018ApJS..235...42A} and \emph{Gaia} DR2 to conclude that the inner halo is dominated by debris from an object that, at infall, was slightly more massive than the Small Magellanic Cloud. The latter authors also argue that the merger must have led to the dynamical heating of the precursor of the Galactic thick disk, approximately 10 Gyr ago.
The dwarf galaxy progenitor of this debris --- now variously called
``Gaia-Enceladus", the ``Gaia Sausage", or collectively the ``Gaia-Enceladus-Sausage'' (GES) --- is thought to have fallen into the Milky Way on a highly eccentric orbit (e $\sim$ 0.85) to account for the predominance of stars with such radial orbits in the inner Galaxy \citep{2018ApJ...863..113H,2018MNRAS.478..611B,2018Natur.563...85H,2019MNRAS.484.4471F}.

\begin{figure}[ht]
\begin{center}
\includegraphics[width=1.\hsize,angle=0]{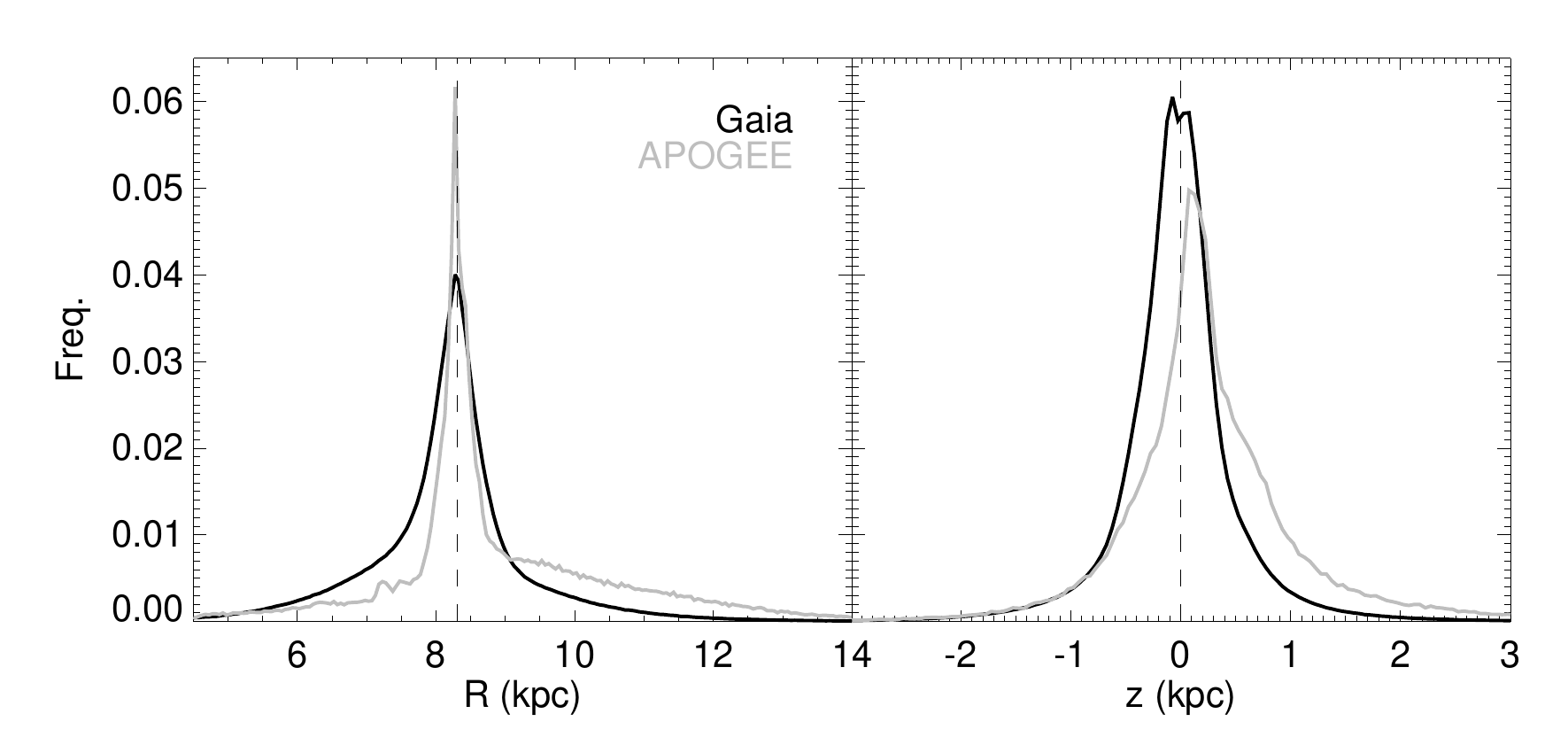}
\end{center}
\caption{({\it Left}) The distribution of distance from the Galactic center projected on the Galactic plane ($R$) in kpc for the APOGEE and \emph{Gaia} sample. The Sun is 
assumed to be situated at $R$ = 8.3 kpc. ({\it Right}) 
The distribution of distance from the Galactic plane for the APOGEE and \emph{Gaia} stars.  The distribution shows that 
most of the 
stars used in this study are in the vertical height range from $-2.0$ $<$ $z$ $<$ 2.0 kpc.  The asymmetry in both panels is a result of the predominance of the Northern Hemisphere observations in the APOGEE sample.}
\label{Gaia_apo_vol}
\end{figure}

\begin{figure*}[ht]
\begin{center}
\includegraphics[width=.497\hsize,angle=0]{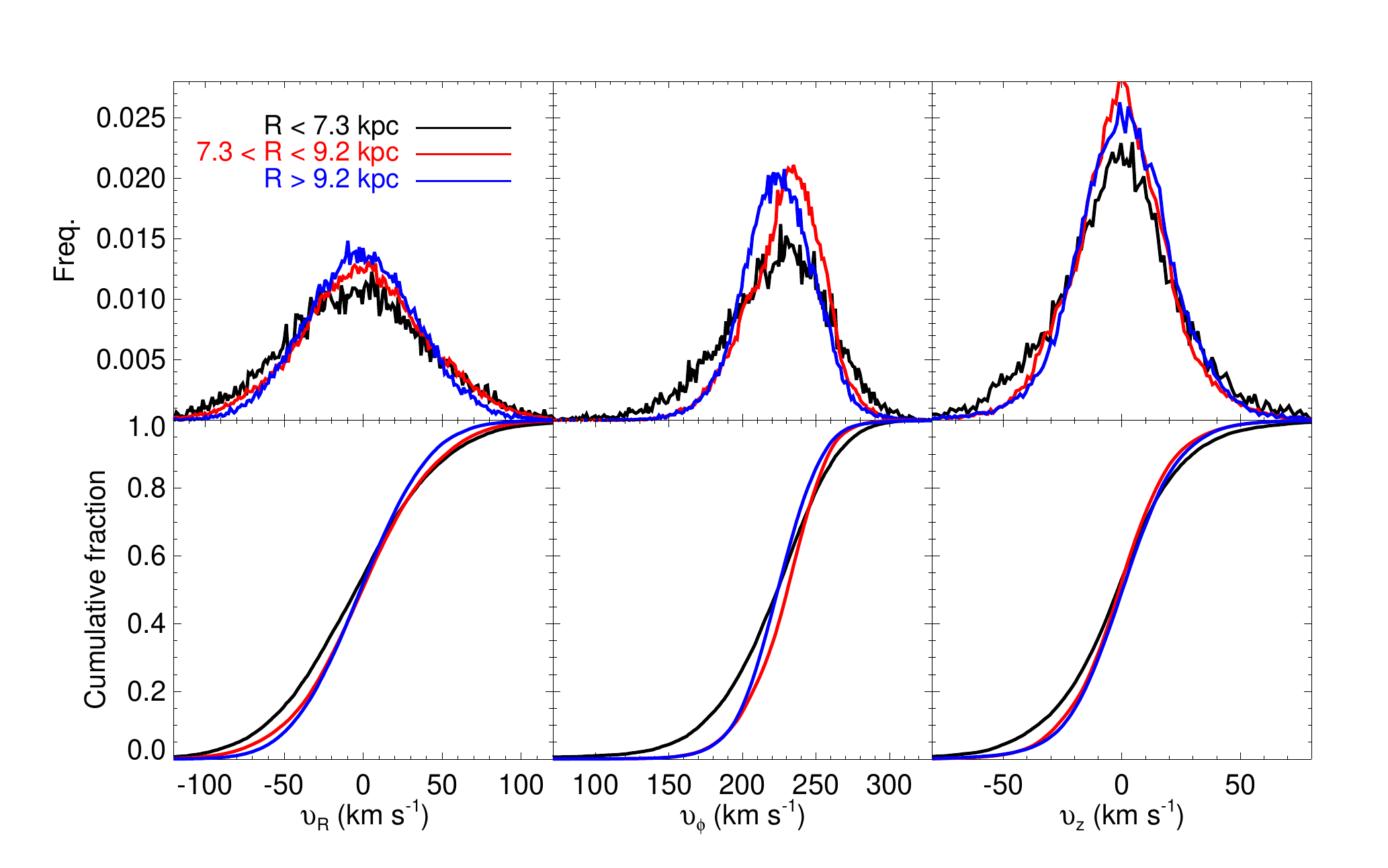}
\includegraphics[width=.497\hsize,angle=0]{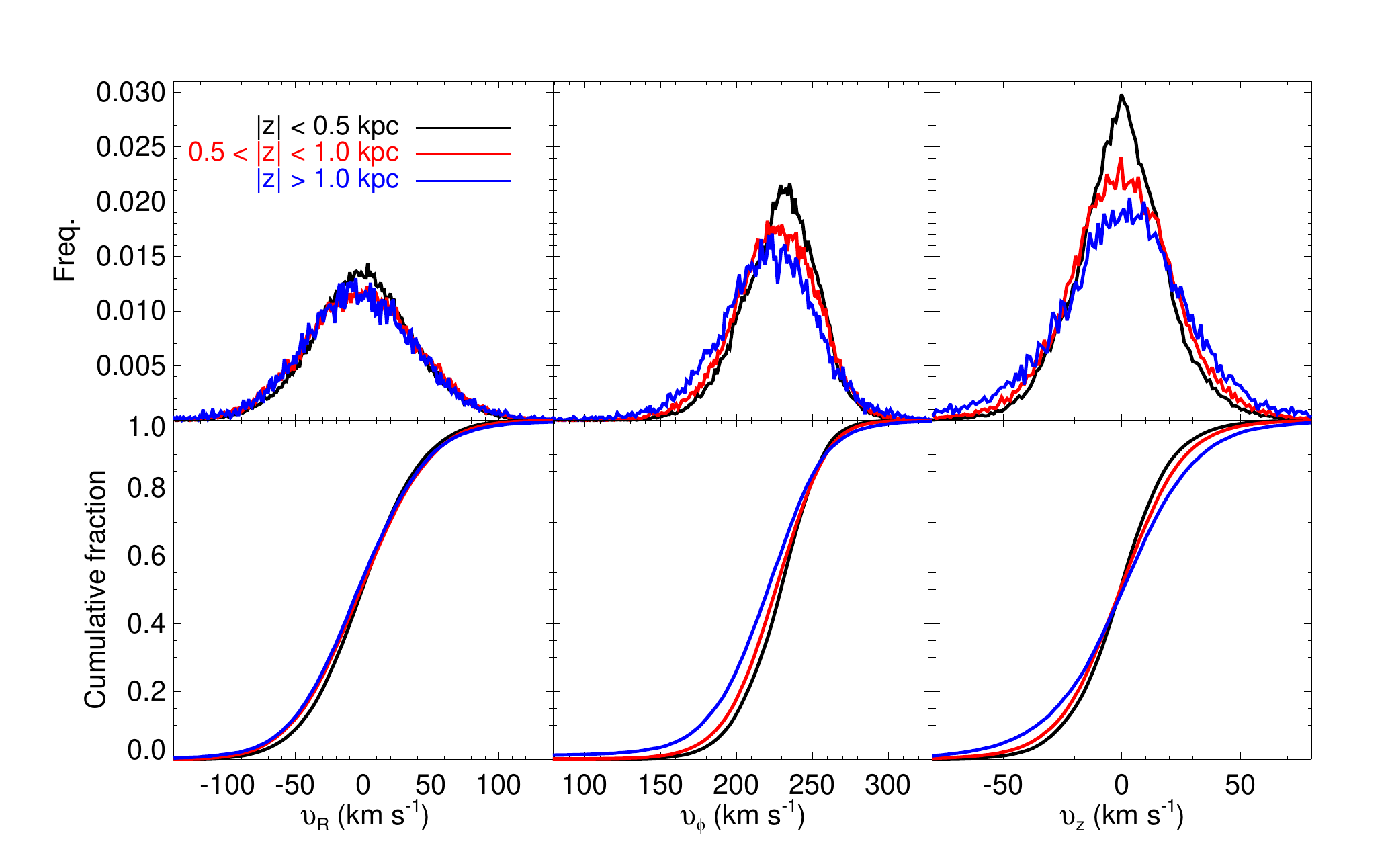}
\includegraphics[width=.497\hsize,angle=0]{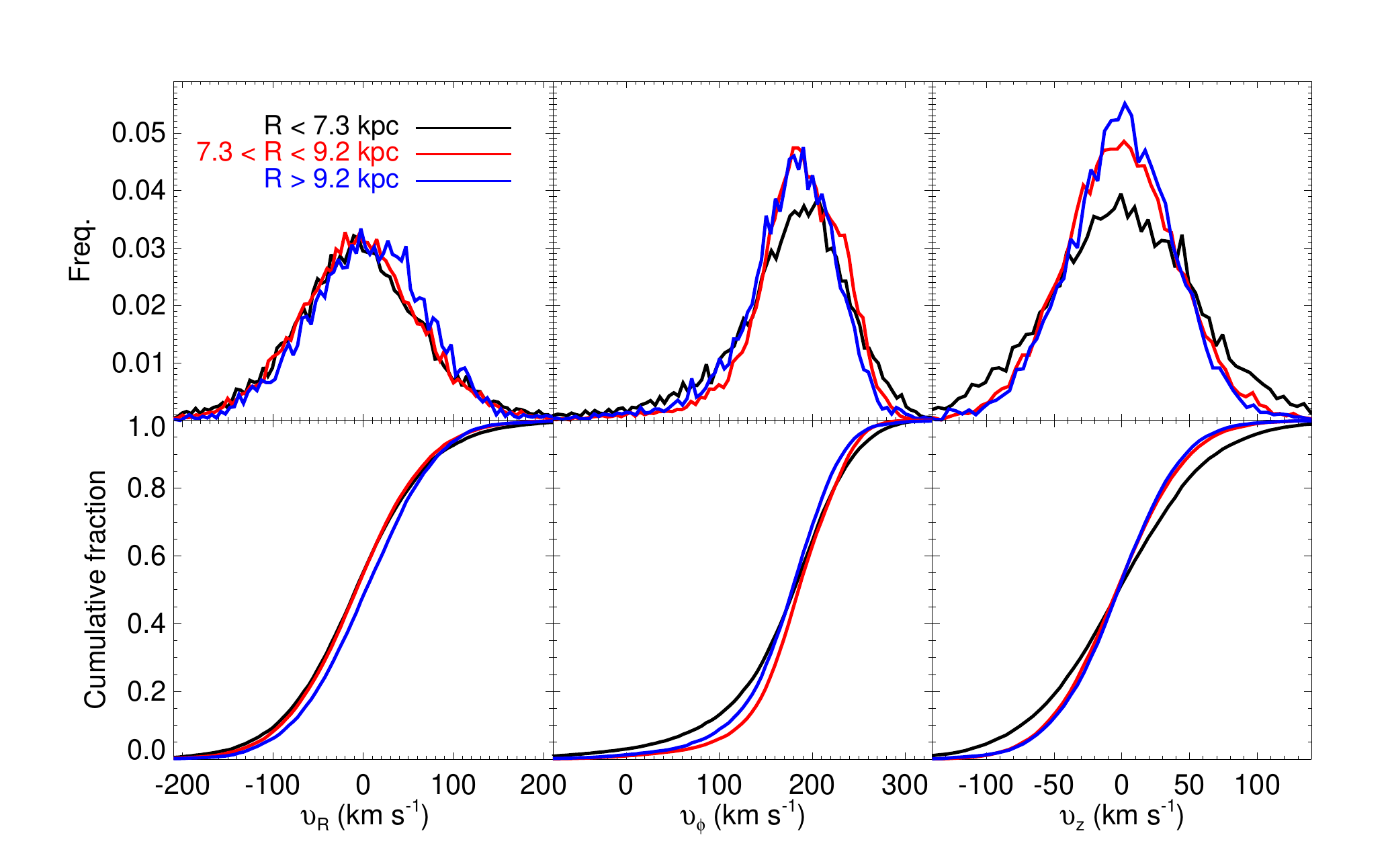}
\includegraphics[width=.497\hsize,angle=0]{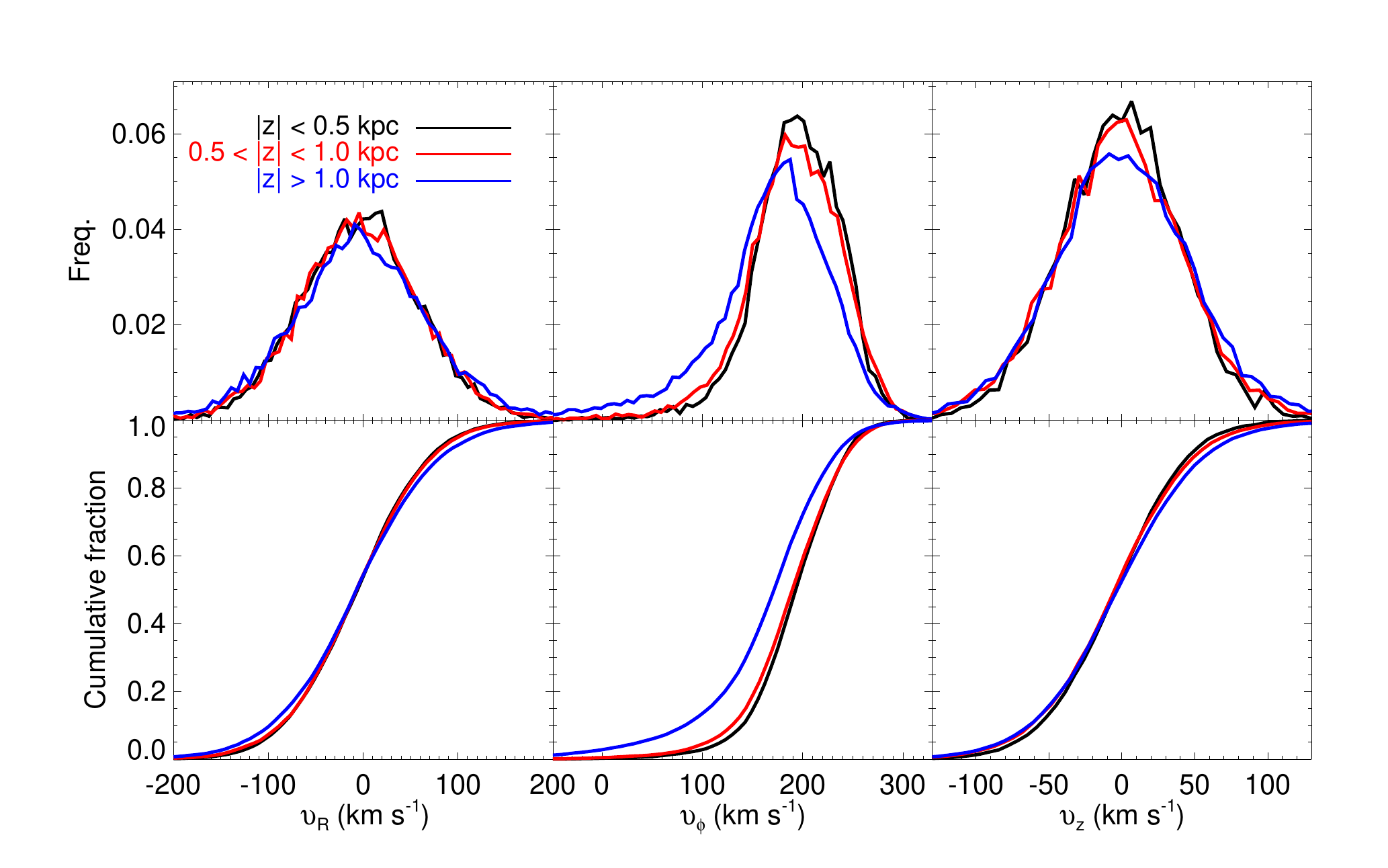}
\includegraphics[width=.497\hsize,angle=0]{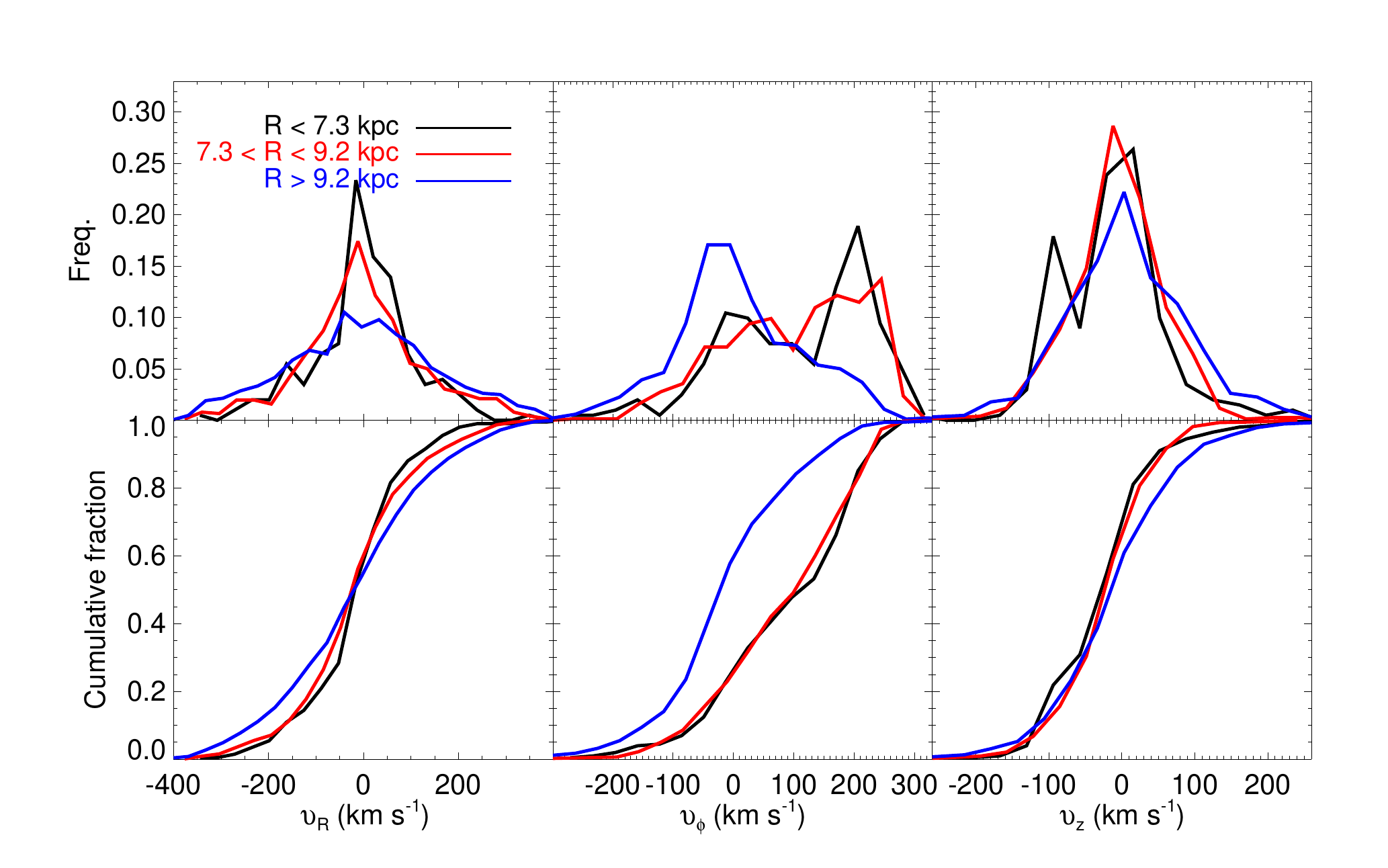}
\includegraphics[width=.497\hsize,angle=0]{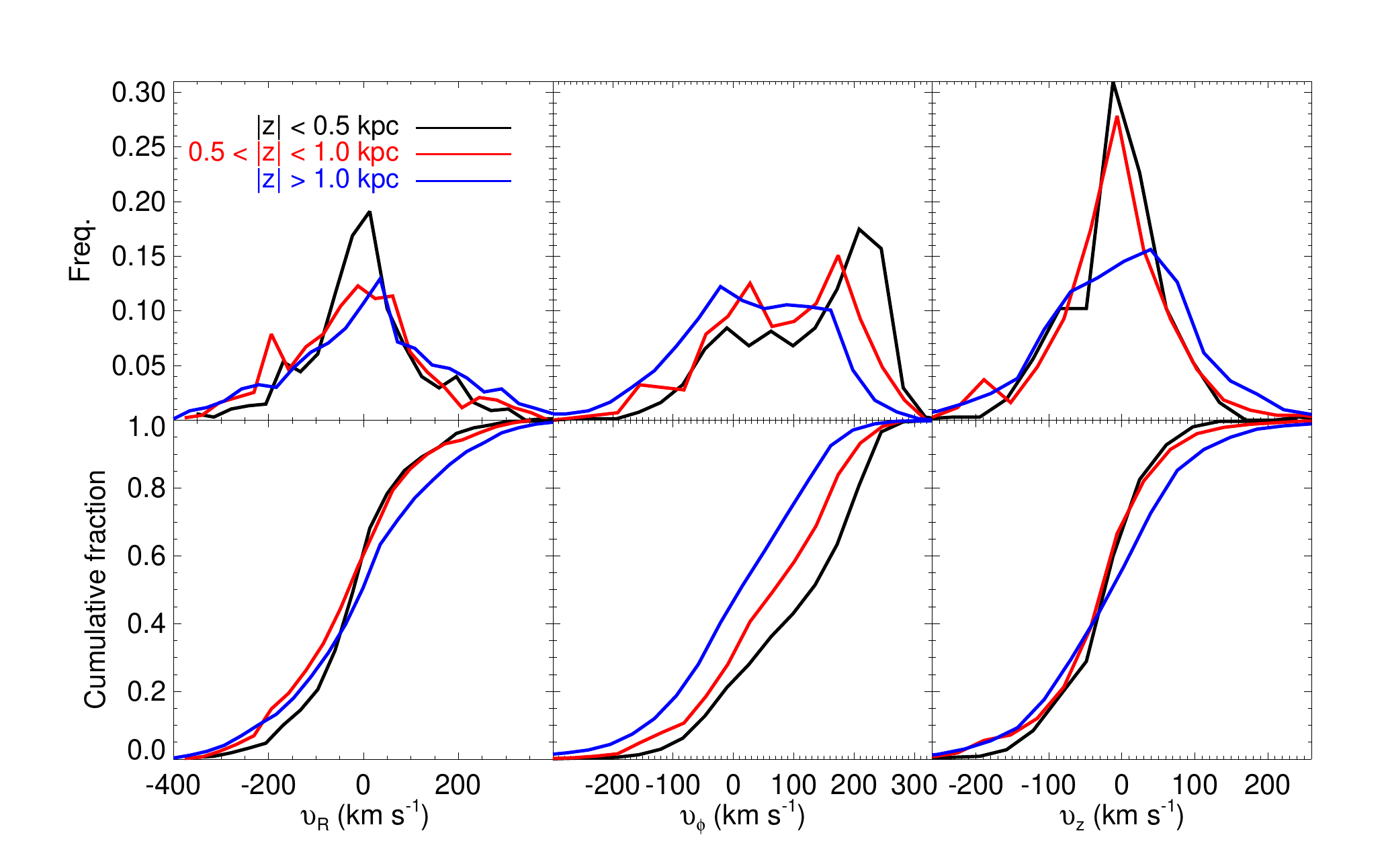}
\end{center}
\caption{Velocity distribution function, together with the cumulative fraction, for the three APOGEE velocity components for different ranges in Galactocentric radius, $R$, and Galactic vertical height, $z$. We show the different spatial ranges in different colors. The top panels represent the chemically selected thin disk, the middle panels the thick disk, and the bottom panels the population with [Fe/H] $<$ $-1.0$, for three different ranges in $R$ (left panels), and $z$ (right panels).}
\label{RZ_distribution}
\end{figure*}

\begin{figure*}[ht]
\begin{center}
\includegraphics[width=1.\hsize,angle=0]{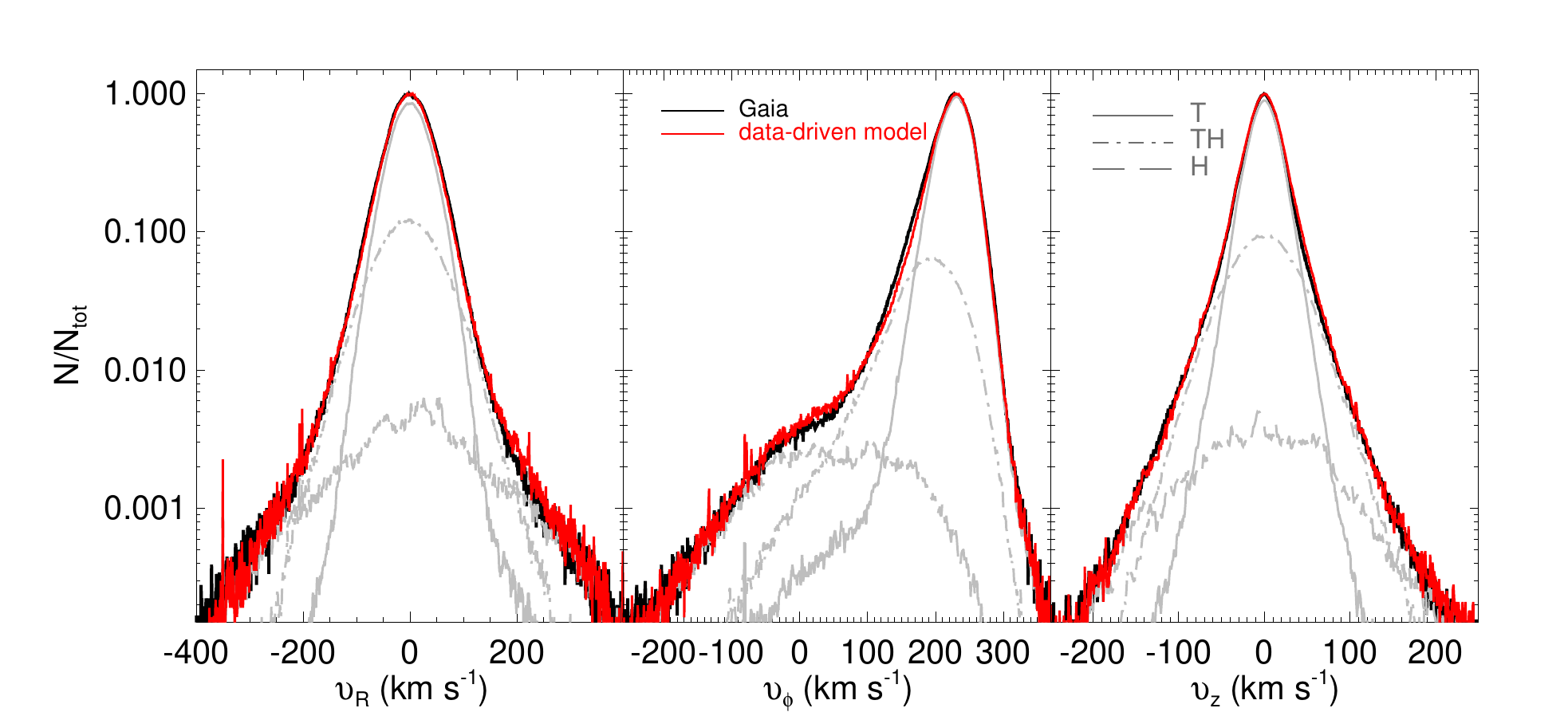}
\end{center}
\caption{The relative velocity distribution functions (normalized to a peak value of unity) for the three velocity components for the \emph{Gaia} DR2 data-set employed in this study ({\it thick black line}) and the
data-driven model generated using APOGEE data for each velocity component ({\it red line}).  
The great similarity of the data and model distributions is clear.
The figure also shows the model decomposition into the thin disk (solid gray line), thick disk (dash dot gray line), and halo (long dash gray line) for the three velocity components.}
\label{data_driven_model}
\end{figure*}

\subsection{The 3D Velocity DF as a Function of $R$ and $z$}
\label{3D_velo}

In a larger context, our chemical-selection approach also allows us to explore the 3D velocity DF as a function of $R$ and $z$ for the different sub-populations.Figure~\ref{Gaia_apo_vol} compares the regions of the sky probed by the APOGEE (grey line) and \emph{Gaia} (black line) samples.
The samples probe a region between 4 $<$ $R$ $<$ 13 kpc, but the majority of stars are located at distances between 7 to 10.5 kpc. The spatial distribution for the vertical height shows that most of the selected stars have $|z|$ $<$ 2.0 kpc (right panel in Figure~\ref{Gaia_apo_vol}). The spatial distributions for the two samples are remarkably similar, despite the fact that the APOGEE survey 
works at brighter magnitudes, but is predominantly targeted at giant stars, whereas the larger \emph{Gaia} sample studied here probes fainter magnitudes, is parallax-error-limited, and is dominated by dwarf stars. Differences in the spatial distributions mainly reflect the fact that \emph{Gaia} is an all-sky survey while the APOGEE DR16 subsample is a pencil-beam survey dominated by Northern Hemisphere observations (where APOGEE has been surveying longer), 
which explains why we see an asymmetry to larger $R$ and an excess for $z$ $>$ 0 with respect to \emph{Gaia} (see right panel in Figure~\ref{Gaia_apo_vol}).


Figure~\ref{RZ_distribution} shows the velocity DF for three different ranges in Galactocentric radius and vertical height for the chemically selected thin disk (top panels), thick disk (middle panels), and halo (bottom panels). We also show the cumulative fraction for each distribution. The different colors show the velocity DFs for three ranges in Galactocentric radius and vertical height. 

In most cases, we do not find large discrepancies between the velocity DF for the three individual components across different ranges of $R$ and $z$, especially for the thin disk and thick disk. The negligible spatial variations in the velocity DF for the different, chemically selected populations justifies and motivates our strategy  
to use the velocity distributions 
discussed in Section~\ref{thin}, \ref{thick}, and \ref{halo_pop} to build a data-driven model of Galactic stellar  populations.

However, it is notable that we observe a small asymmetric-drift variation with $z$ for the thick disk (right figure in the middle panels in Figure~\ref{RZ_distribution}). We also find that the innermost part of the thin disk tends to lag with respect to the rest of the population at larger Galactocentric radii (left figure in the upper panels in Figure~\ref{RZ_distribution}).
On the other hand, for
the [Fe/H] $< -1$ 
group of stars we find strong variations in the contributions of the two components in $\upsilon_{\phi}$ described in Section~\ref{halo_pop}, with 
the thick-disk-like population
more prominent in the inner region ($R < 7.3$ kpc) and closer to the Galactic plane
($|z|$ $<$ 0.5 kpc), and 
the ``inner halo'' population with a non-rotating average motion dominating the distribution for the outer regions in $R$ and $z$ (see bottom panels in Figure~\ref{RZ_distribution}).

\subsection{The Stellar Disk and Halo Contribution in the \emph{Gaia} Sample}
\label{AD_test}

To carry out an unbiased study of the 

spatial distributions
of the Galactic components
for \emph{Gaia} stars, the bulk of which lack abundance information, we use the results presented above to model the contribution from the thin disk, the thick disk, and the halo stars in the much larger \emph{Gaia} sample. The \emph{Gaia} velocity DF observed in the three components of space velocity (see Figure~\ref{vphi_APOGEE}) is a combination of the different components of the stellar disk and the Galactic halo. Using the APOGEE velocity DF measured for the chemically distinguished disk and halo components, we can estimate the fraction of stars that belong to each of these components in the much larger \emph{Gaia} data-set.  	

In the previous section, we verified that the APOGEE subsample is a legitimate proxy for the larger Gaia sample. In particular, we checked which regions of the Galaxy are probed by the \emph{Gaia} sample employed in this exercise as well as the APOGEE sample, and find (Figure~\ref{Gaia_apo_vol}) that they survey comparable volumes.

To assess further the ability of the derived
data-driven model to quantify the number of objects in the \emph{Gaia} sample that are attributable to the thin disk, thick disk, and halo, we apply an Anderson-Darling style statistical test sensitive to discrepancies at low and high values of $\vec{\upsilon}$ \citep{1992nrfa.book.....P}, adopted as follows:


$<S_{G}(\vec{\upsilon}) - S_{A}(\vec{\upsilon})>^{2} /  S_{A}(\vec{\upsilon})(1 - S_{A}(\vec{\upsilon}))$ , \label{equa}


\noindent 
The Anderson-Darling statistic 
is based on the empirical cumulative distribution function.  The cumulative distribution for the \emph{Gaia} sample (S$_{G}(\vec{\upsilon})$) is calculated directly from the individual velocities in the \emph{Gaia} data-set. For the cumulative distribution calculated using the velocity DF from the APOGEE data-set (S$_{A}(\vec{\upsilon})$), we have a mixture of different distributions for the thin disk, thick disk, and halo that we draw directly from the APOGEE data (black distributions in Figures~\ref{velo_thin_apogee}, \ref{velo_thick_apogee}, and \ref{velo_halo_apogee}).
The sum of the three stellar velocity distributions associated with the three Galactic components allows us to derive estimates, 
guided by
Anderson-Darling statistical tests, of the fraction of objects that are part of the thin disk, thick disk, and halo in the {\it Gaia} data set.
We can write $f(\vec{\upsilon})_{gaia} = \sum_{i=1}^{3} \epsilon_{i}   f(\vec{\upsilon})_{apogee}$, where $i$ represents each velocity component, and $\epsilon_{i}$ = ($\epsilon_{thin}$, $\epsilon_{thick}$, $\epsilon_{halo})$ is the fraction of stars associated to that Galactic component. Following the results in Section~\ref{3D_velo}, we assume that the shape of the velocity distribution for each component does not change in the local volume of study; what changes is the fraction of stars in each Galactic component.

 Figure~\ref{data_driven_model} shows the velocity distribution function for the three velocity components. The thick black line is the \emph{Gaia} DR2 data-set employed in this study; red is the data-driven model built from the chemically-selected thin disk, thick disk, and halo. That the thick black {\it Gaia} data lines in Figure ~\ref{data_driven_model} are almost completely obscured by the red model lines is a testament to the veracity of the model to explain the data. For the $\upsilon_{R}$ component, we find the fraction of the Galactic components (thin disk, thick disk and halo, respectively), $\epsilon$ = (0.777, 0.208, 0.015), while for $\upsilon_{\phi}$ we find, $\epsilon$ = (0.881, 0.103, 0.016). Finally, for the vertical component, $\upsilon_{z}$, we obtain $\epsilon$ = (0.799, 0.186, 0.015). Figure~\ref{data_driven_model} also shows the decomposition into the thin disk, thick disk, and halo velocity distribution functions (gray lines) from the data-driven model showing the estimation of the fraction of objects that are part of the different Galactic structures. The radial and the vertical velocity DFs give similar fractions for the three components, however the $\upsilon_{\phi}$ velocity yields a larger number of thin-disk stars compared to the thick disk. It is likely that this difference has to do with the fact that while the distribution of $\upsilon_{R}$ and $\upsilon_{z}$ can be reproduced with a Gaussian function (see Sections~\ref{thin} and \ref{thick}), the distribution of $\upsilon_{\phi}$ is strongly non-Gaussian. The latter is highly skewed to low velocities due to the asymmetric drift, especially for the Galactic thick disk (as seen in Figures~\ref{velo_thick_apogee} and \ref{velo_halo_apogee}), and this non-asymmetric behavior is more challenging for our model to describe. 


Nevertheless, by combining the results calculated using the three individual velocities, and taking the average value, we estimate that 81.9 $\pm$ 3.1 $\%$ of the objects in the selected \emph{Gaia} data-set are thin-disk stars, 16.6 $\pm$ 3.2 $\%$ are thick-disk stars, and 1.5 $\pm$ 0.1 $\%$ belong to the Milky Way halo. 



\section{Thick-disk normalization}
\label{thick_norma}

In this section we determine the local density normalization ($f_{\rho}$ = $\rho_{T}$ /$\rho_{t}$) of the thick disk compared to the thin disk using the velocity distribution function derived above. 
Many previous derivations 
of $f_{\rho}$ were based on 
starcount data derived from photometric parallaxes (e.g., \citealt{1983MNRAS.202.1025G, 2008ApJ...673..864J}),
with the estimates ranging from 1$\%$ to 12$\%$. \cite{2016ARA&A..54..529B} analyzed the results from 25 different photometric surveys conducted since the discovery of the thick disk \citep{1982PASJ...34..365Y} and concluded that $f_{\rho}$ = 4$\%$ $\pm$ 2$\%$ at the solar circle. 
However, starcount approaches to this problem are subject to the degeneracy between the derived scalelengths and scaleheights for each the thin-disk and thick-disk components,
which drives a large uncertainty in $f_{\rho}$ (\citealt{2001ApJ...553..184C, 2002ApJ...578..151S}).
Our novel approach, using the observed {\it velocity} distribution functions for the thin disk, thick disk and halo from APOGEE as a data driven-model applied to the \emph{Gaia} data-set, is not affected by this degeneracy.
Moreover, we are able to  analyze the behavior of $f_{\rho}$ in the $R$-$z$ plane, averaged
over the Galactocentric polar angle, $\phi$.

\begin{figure}[ht]
\begin{center}
\includegraphics[width=1.\hsize,angle=0]{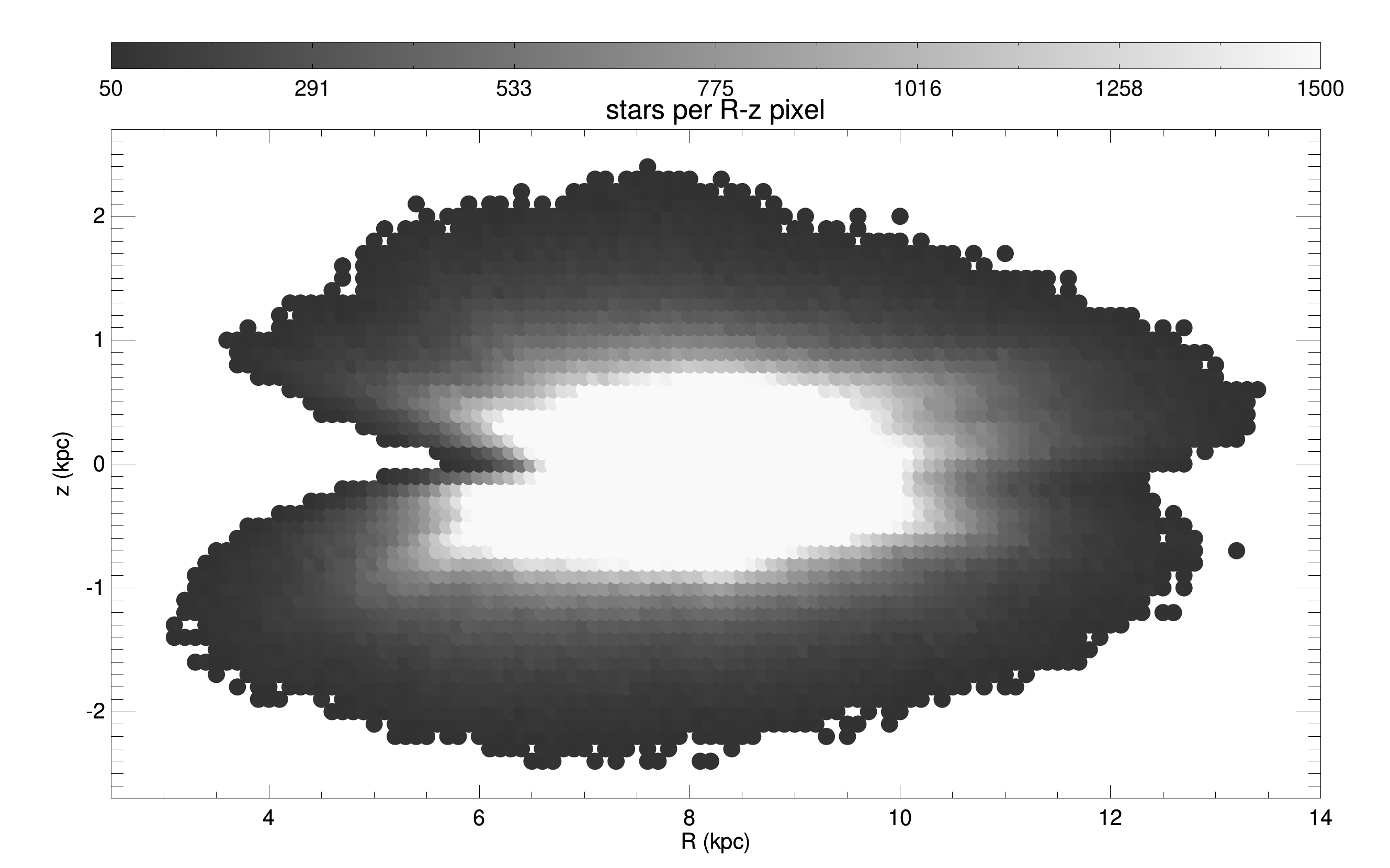}
\end{center}
\caption{Stellar number density as a function of Galactic cylindrical coordinates $R$ and $z$. The density is shown on a linear scale and coded from black to white. The larger number of stars in the \emph{Gaia} data-set per $R$-$z$ pixel are within $-1$ $<$ $z$ $<$ 1 kpc and 6 $<$ $R$ $<$ 10 kpc.}
\label{stellar_count}
\end{figure}

\begin{figure}[ht]
\begin{center}
\includegraphics[width=1.\hsize,angle=0]{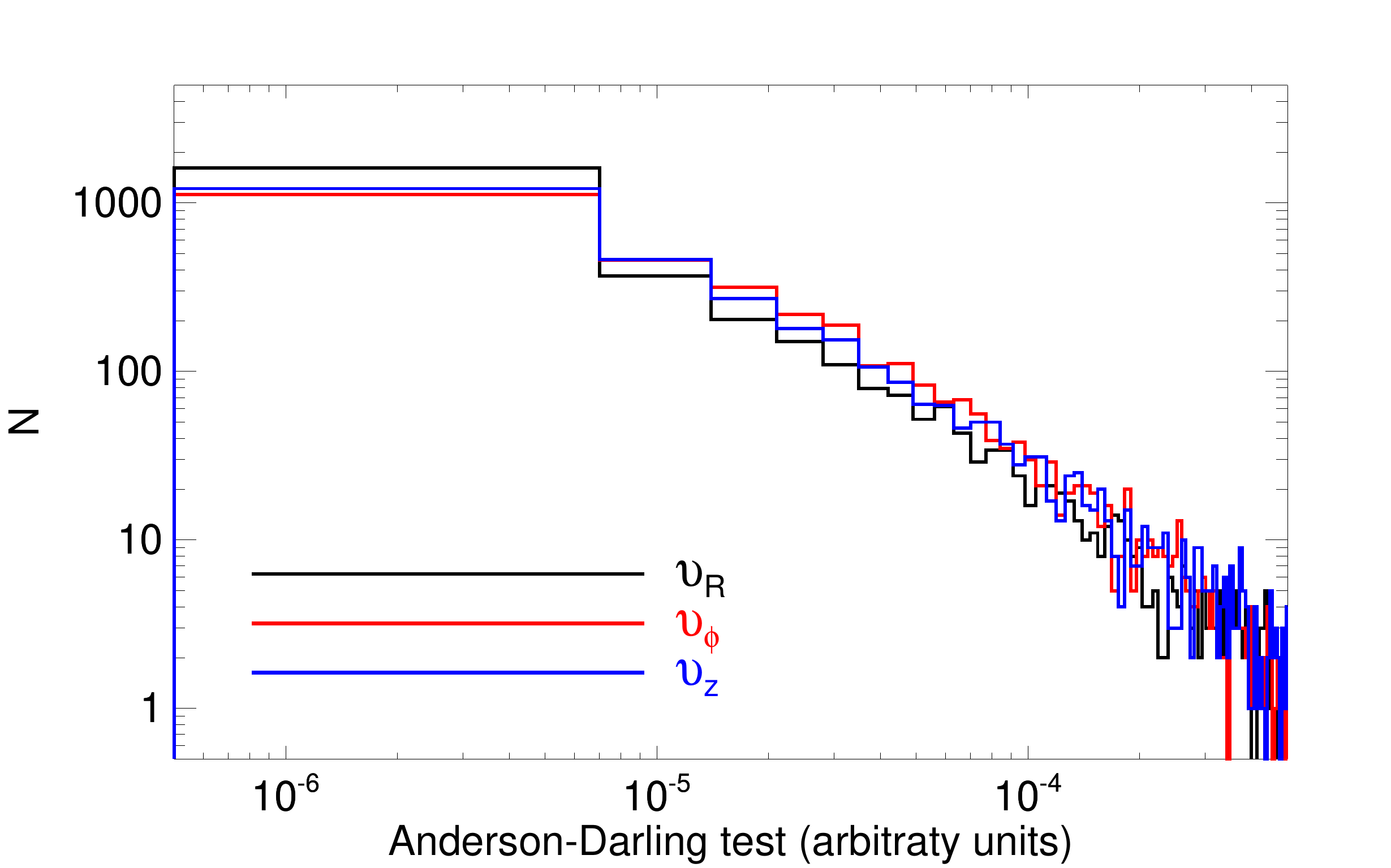}
\includegraphics[width=1.\hsize,angle=0]{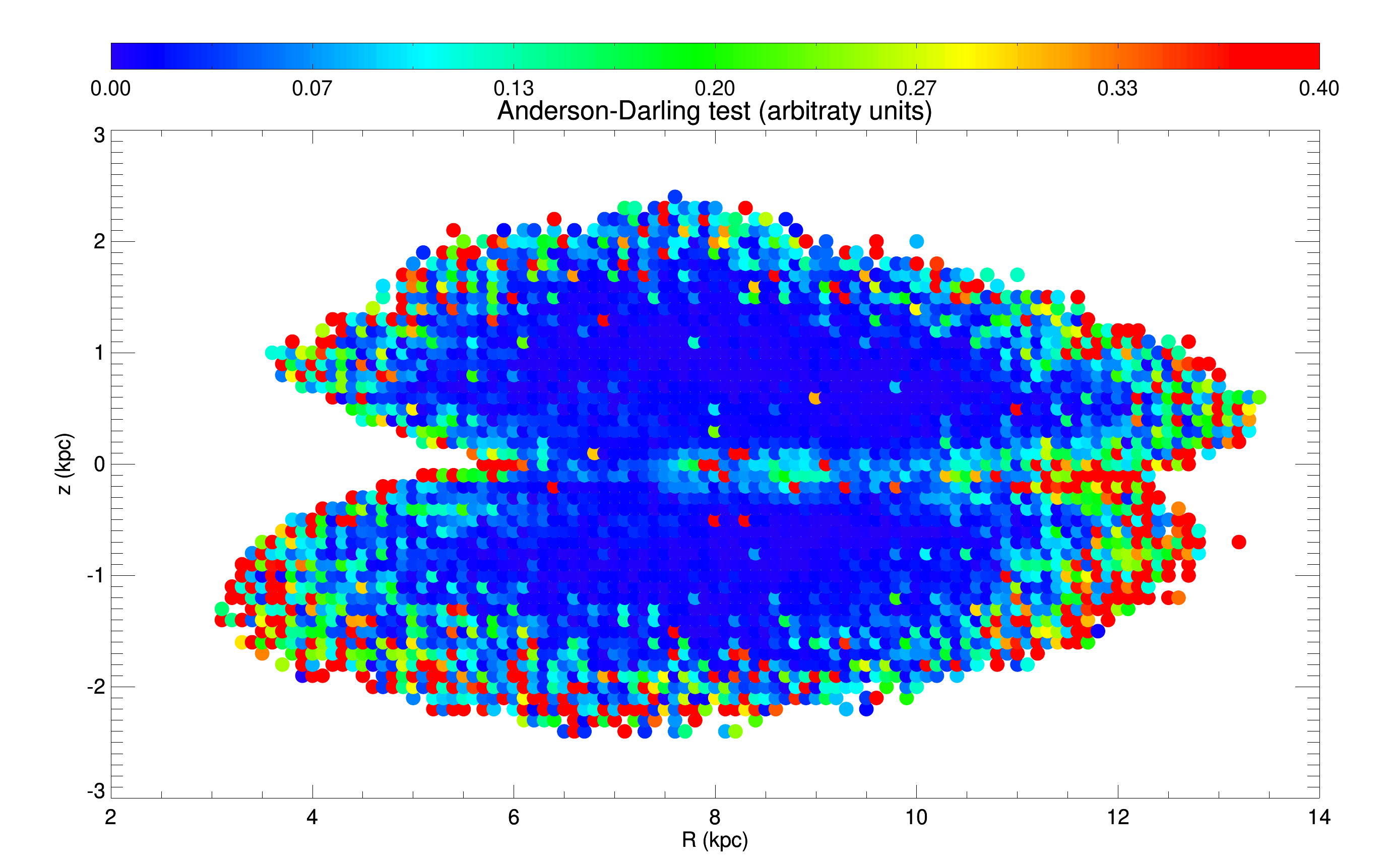}
\end{center}
\caption{({\it Top}) Values from the Anderson-Darling statistic as a goodness-of-fit test for the individual space velocities. The values can be interpreted as the maximum deviation between the cumulative distribution of the data and the data-driven model function. ({\it Bottom}) The Anderson-Darling test values for the three velocity components added in quadrature
in the $R$-$z$ plane.
}
\label{AF}
\end{figure}

To build a given $R$-$z$ pixel, we create intervals of 0.1 kpc in position,
and use only pixels where the number of stars for a given interval is N($R_{i}$,$z_{i}$) $\geq$ 50.
Figure~\ref{stellar_count} shows the stellar number density as a function of Galactic cylindrical coordinates $R$ and $z$ created through this exercise.
The largest number of stars per pixel in the magnitude-limited \emph{Gaia} sample are within $-1$ $<$ $z$ $<$ 1 kpc and 6 $<$ $R$ $<$ 10 kpc. Using the data-driven model described in Section~\ref{model}, we create different velocity DFs, where our free parameters are the fraction of thin-disk, thick-disk, and halo stars in intervals of 1$\%$ for each population. For a given $R$-$z$ pixel, we use the observed velocity DF in the area of study, and we perform an Anderson-Darling style statistical test (as was done in Sec.~\ref{AD_test}). The minimum value in the statistical test (i.e., the
maximum deviation between the cumulative distribution of the data and the data-driven model function) is used to estimate the fraction, and hence the density normalization, of each population. We perform this analysis for the three individual velocity components separately. Figure~\ref{AF} shows the results from the statistical Anderson-Darling test for the three velocities (top panel). We find that the distributions are very similar regardless of the velocity components used. The bottom panel of Figure~\ref{AF} also shows 
the Anderson-Darling test values for the three individual velocities added in quadrature in the $R$-$z$ plane. $R$-$z$ pixels with lower numbers of stars tend to have larger Anderson-Darling test values, i.e., a larger deviation between the data and the model.

\begin{figure*}[ht]
\begin{center}
\includegraphics[width=.49\hsize,angle=0]{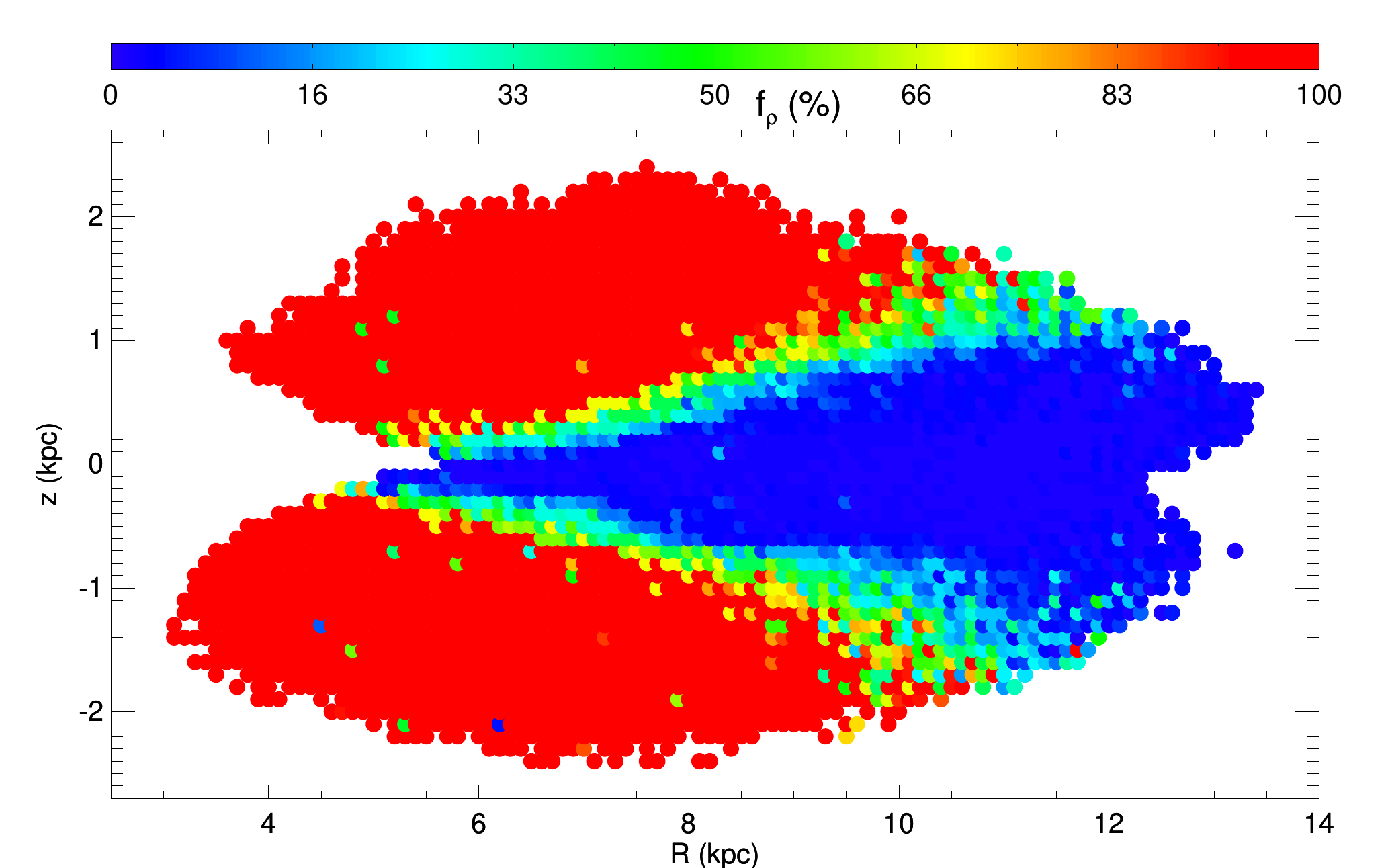}
\includegraphics[width=.49\hsize,angle=0]{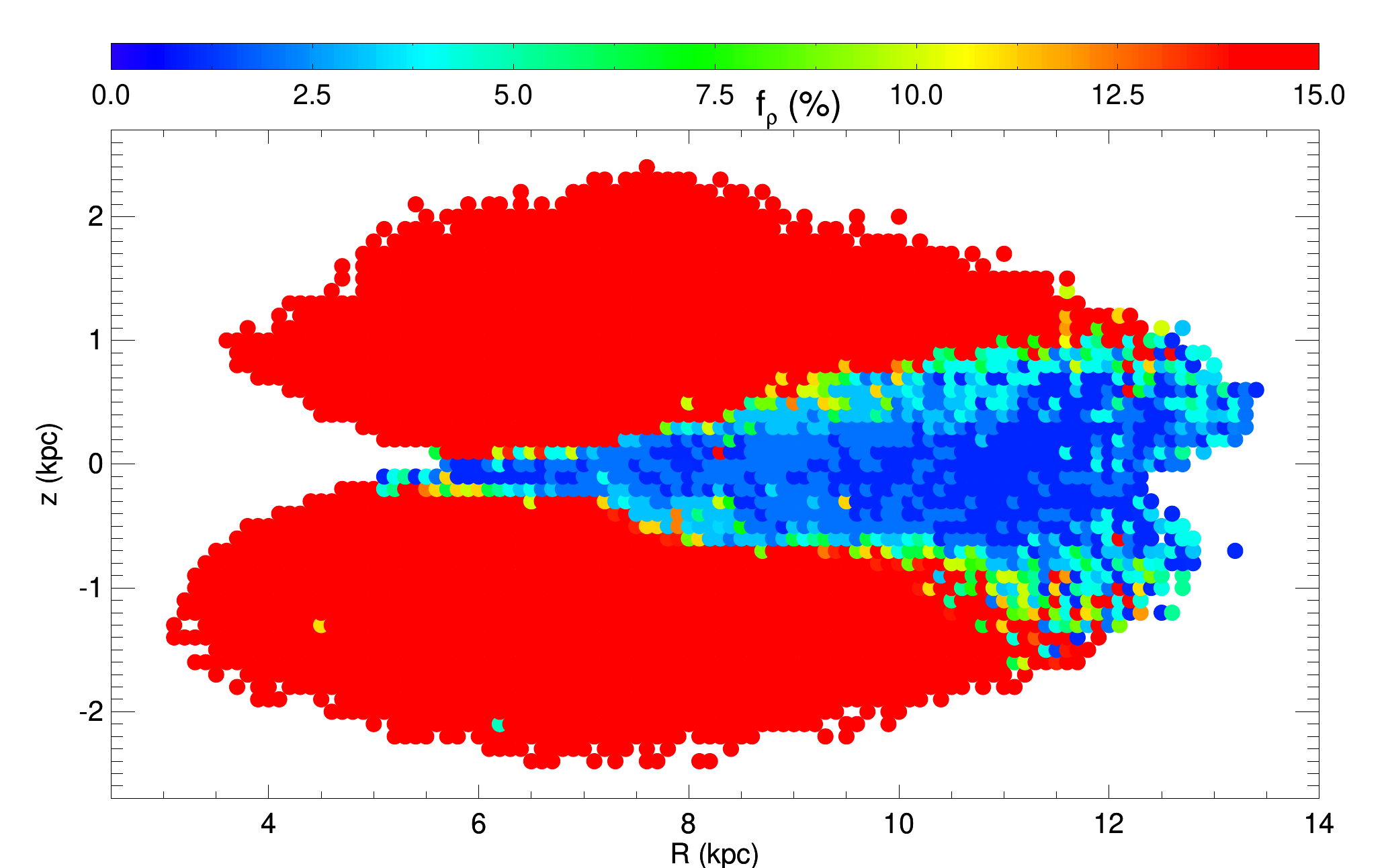}
\end{center}
\caption{The relative density of \emph{Gaia} stars in the $R$-$z$ plane color-coded by $f_{\rho}$ the fraction of stars that belong to the thick disk versus the thin disk ($f_{\rho}$).
The dark blue regions, where the $f_{\rho}$ values range from nearly zero to $\sim$ 15 $\%$, are those dominated by the thin disk population.  Green regions indicate a \emph{transition region} where we have a more even mix between thin disk and thick disk populations (i.e., 
$f_{\rho}$ $\sim$ 50 $\%$. Finally, the red areas are where the 
Galactic thick disk dominates ($f_{\rho}$ $>$ 85 $\%$). The right panel is the same as the left panel, but with a different color-coding that highlights differences at lower thin disk fractions. The visualization of this new range in $f_{\rho}$ allows us to see smaller differences in the disk normalization for a given position.}
\label{density_thin_RZ}
\end{figure*}

Once we know the fraction of stars that belong to the thin disk, thick disk, and halo, calculation of the thick-disk normalization, $f_{\rho}$, in the $R$-$z$ plane is straightforward. For the final value, we average between the $f_{\rho}$ given for each velocity component. We present the results in Figure~\ref{density_thin_RZ}, where
the $R$-$z$ plane is color-coded by $f_{\rho}$. These figures, for the first time, reveal in detail how $f_{\rho}$ varies across the Galaxy.  

From inspection of Figure \ref{density_thin_RZ}, one can clearly see where the thin-disk population dominates (dark blue), where the $f_{\rho}$ values range from nearly zero to $\sim$ 15 $\%$. We also see a \emph{transition region}, where we have a more even mix between thin-disk and thick-disk populations (green), indicated by where the maps have $f_{\rho}$ $\sim$ 50 $\%$. Finally, we also observe the region where the population that belongs to the Galactic thick disk dominates (red; see right panel in Figure~\ref{density_thin_RZ}), shown by where $f_{\rho}$ $>$ 85 $\%$. Interestingly, we observe that the density ratio in the \emph{transition region} shows differences between the North and South. The area around
$R\sim 10$ kpc does not show the same transition in density at $z$ $\sim$ $-1.5$ kpc as at $z$ $\sim$ 1.5 kpc. This could be an effect of the Galactic warp (e.g., \citealt{2006ApJ...643..881L,2019A&A...627A.150R} and references therein), but also this could be related to excitation of wave-like structures creating a wobbly galaxy \citep{2012ApJ...750L..41W,2013MNRAS.436..101W}. Furthermore, Figure~\ref{density_thin_RZ} shows a radial trend, in the sense that at increasing $R$ the thin disk dominates at larger $z$ (i.e., the blue region in Figure~\ref{density_thin_RZ} gets thicker at larger Galactocentric radius). This phenomena could be interpreted as the thin disk flaring (e.g., \citealt{2014Natur.509..342F,2019MNRAS.483.3119T}), but also can be related to the different scalelength of the thin disk with respect to the thick disk, where the scalelength for the thick disk is shortened with respect to the one for the thin disk. The latter suggestion is in line with the recent literature on this debated topic. For example, \cite{2011ApJ...735L..46B} and \cite{2012ApJ...753..148B} suggested a shorter scalelength for the chemically distinguished high--$\alpha$ disk compared to the low--$\alpha$ disk. 

Moreover, in the right panel of Figure~\ref{density_thin_RZ}, we have the same \emph{Gaia} data $R$-$z$ plane color-code by $f_{\rho}$, but in this case the figure is restricted from 0 to 15$\%$. The visualization of this new range in $f_{\rho}$ allows us to see smaller differences in the disk normalization for a given position. We suggest that vertical oscillations in the disk (e.g., \citealt{2018MNRAS.480.4244C}) may be responsible for these small fluctuations in $f_{\rho}$ for the thin Galactic disk, by introducing
small changes in the shape of the velocity DF in such a dynamically active disk. 

In the end, using our novel approach based on {\it velocities}, rather than starcounts,
we find the local thick-to-thin disk density normalization to be $\rho_{T}(R_{\odot})$/$\rho_{t}(R_{\odot})$ = 2.1 $\pm$ 0.2 $\%$. 

\begin{figure}[ht]
\begin{center}
\includegraphics[width=1.\hsize,angle=0]{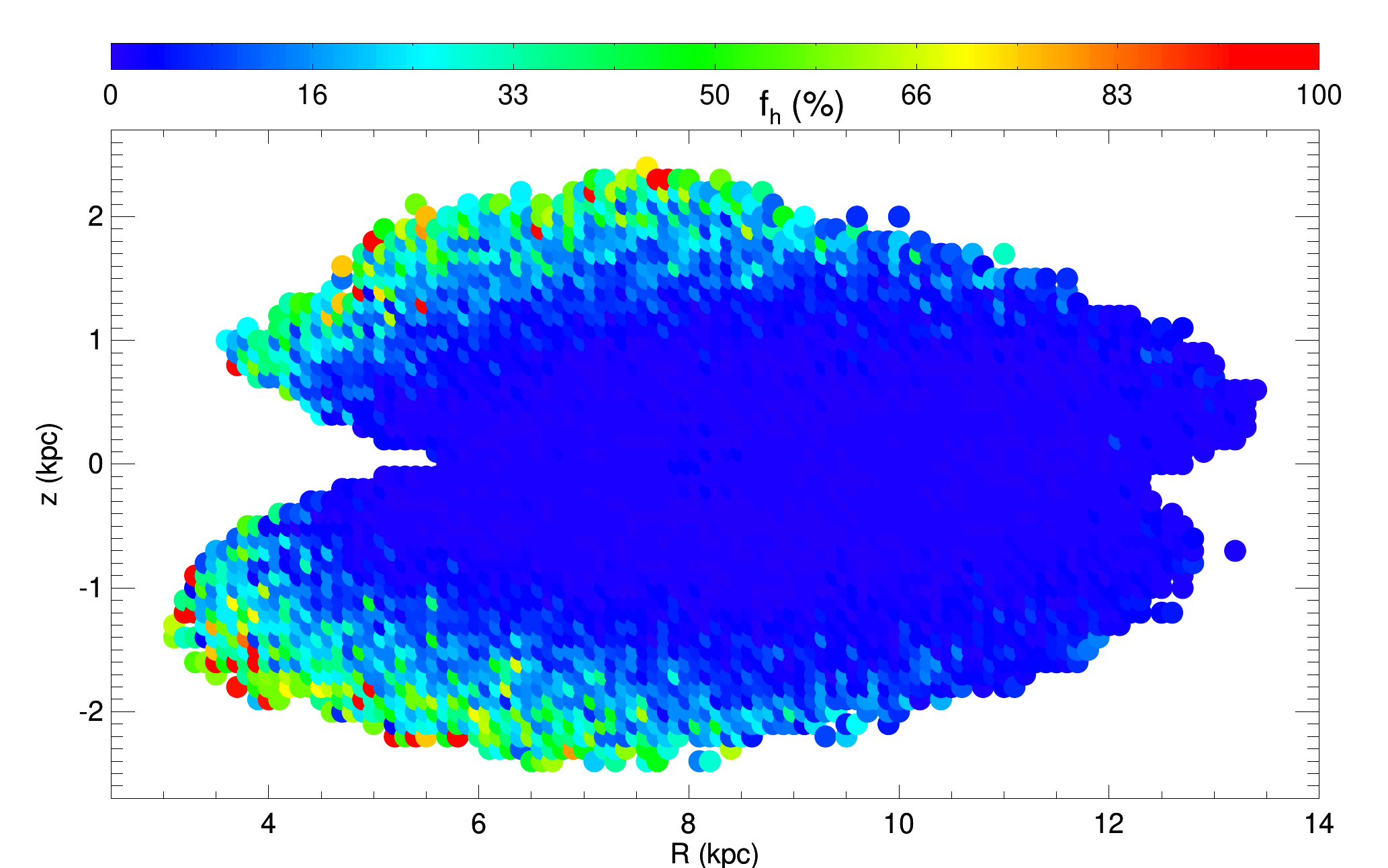}
\end{center}
\caption{The same as Figure \ref{density_thin_RZ}, but
color-coded by the ratio of stars that belong to the halo compared to those belonging to the disk (thin + thick).  
The areas where the halo is most dominant
occur at $R$ $<$ 8 kpc and $|z|$ $>$ 1.3 kpc.}
\label{halo_disk}
\end{figure}

Figure~\ref{halo_disk} shows the halo-to-disk density normalization. We find that the halo is most dominant 
in the \emph{Gaia} data-set in the region $R$ $<$ 8 kpc and $|z|$ $>$ 1.3 kpc. We find the local halo-to-disk density normalization to be $\rho_{H}(R_{\odot})$/($\rho_{T}(R_{\odot})$ + $\rho_{t}(R_{\odot})$) = 1.2 $\pm$ 0.6 $\%$. 


\section{Summary and Conclusions}
\label{Conclusion}

Combining the precise stellar abundances from the APOGEE survey with the astrometry from \emph{Gaia}, we study the velocity DF for chemically selected low-$\alpha$ (thin-disk), high-$\alpha$ (thick-disk), and halo ([Fe/H] $< -1$) stars. Using the kinematical properties of these sub-samples, we built a data-driven model,
and used it to dissect
a 20\% parallax-error-limited \emph{Gaia} sample to understand the contribution of the different Galactic structural components to the velocity-space DF as a function of Galactic cylindrical coordinates $R$ and $z$. We find that 81.9 $\pm$ 3.1 $\%$ of the objects in the selected \emph{Gaia} data-set are thin-disk stars, 16.6 $\pm$ 3.2 $\%$ are thick-disk stars, and 1.5 $\pm$ 0.1 $\%$ belong to the Milky Way halo. 

The local fraction of the Milky Way thick disk, $\rho_{T}(R_{\odot})$/$\rho_{t}(R_{\odot})$, is still under debate, as evidenced by the large spread in derived values
--- ranging from 2$\%$ \citep{1983MNRAS.202.1025G} to 12$\%$ \citep{2008ApJ...673..864J} --- 
from starcounting methods, which are notoriously fraught with model degeneracies.
Our analysis, based on the velocity characteristics of chemically selected populations, helps to break the aforementioned degeneracies, and favors lower values for the normalization (e.g., \citealt{2012ApJ...753..148B}). 
We find the local thick-to-thin disk density normalization to be $\rho_{T}(R_{\odot})$/$\rho_{t}(R_{\odot})$ = 2.1 $\pm$ 0.2 $\%$, a result consistent with the lower end of values derived using starcount/density analyses, but determined in a completely different way.

Using the same methodology, the local halo-to-disk density normalization is found to be $\rho_{H}(R_{\odot})$/($\rho_{T}(R_{\odot})$ + $\rho_{t}(R_{\odot})$) = 1.2 $\pm$ 0.6 $\%$. 

This is several times larger than other values found recently using kinematically selected samples.  For example,  \cite{2020MNRAS.tmp...78A}, using high tangential
velocity stars ($\upsilon_{t}$ $>$ 200 km s$^{-1}$) in \emph{Gaia} DR2, found the local halo-to-disk density normalization to be 0.47$\%$. This is in agreement with the value (0.45$\%$) found in \cite{2018A&A...615A..70P}, where the local stellar halo was selected through Action Angle distributions using \emph{Gaia} DR1 and RAVE.  The differences between these kinematics-based halo samples and our chemically selected halo ([Fe/H] $<$ $-1.0$)
might be explained by our halo sample including a non-negligible fraction of metal-poor, but
kinematically colder stars that may be related to the thick disk.
Their inclusion may inordinantly elevate our derived  halo-to-disk density normalization.

While overall a chemical discrimination of stellar populations, as undertaken here, does a remarkably good job, and provides many benefits over other population-separation methods, ultimately our approach is limited by the degree to which the metal-weak thick disk and halo populations overlap in chemical-abundance spaces like that shown in Figure 1. 
In the future, this chemical overlap may be overcome by either the use of a larger number of chemical dimensions, or the combined use of kinematics and chemistry to overcome this one area of population overlap.

\begin{acknowledgements}

BA acknowledge helpful conversations with B. K. Gibson, K. C. Freeman, I. Minchev and A. C. Robin, as well as comments from the anonymous referee that improved the manuscript.  BA, SRM, CRH, and XC appreciate funding from NSF grant AST-1909497.
T.C.B. acknowledges partial support from grant PHY 14-30152  (Physics Frontier Center / JINA-CEE), awarded by the U.S. National Science Foundation. Funding for the Sloan Digital Sky Survey IV has been provided by the Alfred P. Sloan Foundation, the U.S. Department of Energy Office of Science, and the Participating Institutions. SDSS-IV acknowledges support and resources from the Center for High-Performance Computing at the University of Utah. The SDSS web site is www.sdss.org.

SDSS is managed by the Astrophysical Research Consortium for the Participating Institutions of the SDSS Collaboration including the Brazilian Participation Group, the Carnegie Institution for Science, Carnegie Mellon University, the Chilean Participation Group, the French Participation Group, Harvard-Smithsonian Center for Astrophysics, Instituto de Astrof\'isica de Canarias, The Johns Hopkins University, Kavli Institute for the Physics and Mathematics of the Universe (IPMU) / University of Tokyo, Lawrence Berkeley National Laboratory, Leibniz Institut f\"ur Astrophysik Potsdam (AIP), Max-Planck-Institut f\"ur Astronomie (MPIA Heidelberg), Max-Planck-Institut f\"ur Astrophysik (MPA Garching), Max-Planck-Institut f\"ur Extraterrestrische Physik (MPE), National Astronomical Observatories of China, New Mexico State University, New York University, University of Notre Dame, Observat\'orio Nacional / MCTI, The Ohio State University, Pennsylvania State University, Shanghai Astronomical Observatory, United Kingdom Participation Group, Universidad Nacional Aut\'onoma de M\'exico, University of Arizona, University of Colorado Boulder, University of Oxford, University of Portsmouth, University of Utah, University of Virginia, University of Washington, University of Wisconsin, Vanderbilt University, and Yale University.

This publication made use of NASA's Astrophysics Data System. This work has made use of data from the European Space Agency (ESA) mission {\it Gaia} (\url{https://www.cosmos.esa.int/gaia}), processed by the {\it Gaia} Data Processing and Analysis Consortium (DPAC, \url{https://www.cosmos.esa.int/web/gaia/dpac/consortium}). Funding for the DPAC has been provided by national institutions, in particular the institutions participating in the {\it Gaia} Multilateral Agreement. This research has made use of the Spanish Virtual Observatory (http://svo.cab.inta-csic.es) supported from the Spanish MINECO/FEDER through grant AyA2014-55216

\end{acknowledgements}

\end{document}